\documentclass[aps,twocolumn,floatfix,superscriptaddress]{revtex4-1}
\usepackage{graphicx}
\usepackage{dcolumn}
\usepackage{bm}
\usepackage{amssymb}
\usepackage{amsmath}
\usepackage{subfigure}
\usepackage{physics}
\usepackage{sublabel}
\usepackage[utf8]{inputenc}
\renewcommand{\arraystretch}{1.7}
\usepackage{hyperref}
\usepackage{mathtools}
\usepackage{float}
\begin{document}

\title{Ballistic graphene array for ultra-high pressure sensing}

\author{Abhinaba Sinha}
\affiliation{Department of Electrical Engineering, Indian Institute of Technpology Bombay, Powai, Mumbai-400076, India}
	
\author{Pankaj Priyadarshi}
\affiliation{School of Engineering, University of Warwick, Coventry- CV47AL, United Kingdom}
	
\author{Bhaskaran Muralidharan}
\email[E-mail:~]{bm@ee.iitb.ac.in}
\affiliation{Department of Electrical Engineering, Indian Institute of Technpology Bombay, Powai, Mumbai-400076, India}
	
\begin{abstract}
Atomically thin two-dimensional materials such as graphene exhibit extremely high-pressure sensitivity compared to the commercially used pressure sensors due to their high surface-to-volume ratio and excellent mechanical properties. The smaller piezoresistance of graphene across different transport regimes limits its pressure sensitivity compared to other two-dimensional materials. Using membrane theory and thin-film adhesivity model, we show miniaturization as means to enhance the overall performance of graphene pressure sensors. Our findings reveal that ballistic graphene can be configured to measure ultra-high pressure ($\approx 10^{9}$ Pa) with many-fold higher sensitivity per unit area than quasi-ballistic graphene, diffusive graphene, and thin layers of transition metal dichalcogenides. Based on these findings, we propose an array of ballistic graphene sensors with extremely high-pressure sensitivity and ultra high-pressure range that will find applications in next-generation NEMS pressure sensors. The performance parameters of the array sensors can be further enhanced by reducing the size of graphene membranes and increasing the number of sensors in the array. The methodology developed in this paper can be used to explore similar applications using other two-dimensional materials. 
\end{abstract}
	
\maketitle
\section{Introduction}
The discovery of piezoresistance in silicon and germanium in 1954 laid the foundation for silicon strain gauges~\cite{Smith1954}. Following this development, the first silicon diaphragm-based pressure sensor was commercially introduced in 1958~\cite{Bryzek1990}. Further progress in silicon fabrication processes such as anisotropic etching~\cite{Kendall1975, Bean1978, Bassous1978}, ion implantation~\cite{Gandhi1983}, anodic bonding~\cite{Pomerantz1968, Wallis1969}, and micro-machining processes~\cite{Bryzek1990} paved the way for a reduction in thickness of the membranes and the sensor dimensions. The entire class of sensors that came into being as a result of these developments is known as thin-film micro-electro-mechanical (MEMS) pressure sensors. \\
\indent Thin-film MEMS pressure sensors are extremely useful for pressure sensing due to their high sensitivity and compact size~\cite{Gong2001,Smith2013}. The mathematical expression for pressure sensitivity (PS) of silicon obtained by Gong~\textit{et al.}~\cite{Gong2001} predicts an increase in the PS with the reduction in membrane thickness. Consequently, atomically thin 2-D materials are expected to have very high PS and are considered as suitable replacements for conventional membranes in pressure sensors. This fact was experimentally confirmed by Smith~\textit{et. al}~\cite{Smith2013} and Wagner~\textit{et. al}~\cite{Wagner2018} on graphene and $\mathrm{PtSe_{2}}$ respectively. \\
\indent Usually, 2-D Dirac materials such as graphene have a lower gauge factor (GF) than non-Dirac 2-D materials such as layered transition metal dichalcogenides (TMDs), phosphorene, arsenene, and to name a few due to the presence of robust Dirac cones~\cite{Sinha2019, Sinha2020}. Hence, non-Dirac 2-D materials have higher PS than 2-D Dirac materials~\cite{Zhang2017,Nourbakhsh2018,Manzeli2015,Hosseini2015,Zhang2021,Wagner2018,An2019}. Apart from PS, the performance of a membrane pressure sensor is determined by its yield pressure ($P_{yp}$, pressure at the yield point)~\cite{Gong2001}, and critical pressure ($P_{cr}$, the pressure required to delaminate a membrane from the substrate)~\cite{Koenig2011}. Amongst different 2-D materials, graphene is considered a strong contender for next-generation pressure sensors because of its high elastic limit (more than $20\%$)~\cite{Lee2008}, high adhesivity~\cite{Koenig2011}, and high impermeability~\cite{Bunch2008}. Despite the excellent overall properties of graphene, recent studies have shown many-fold higher PS of $\mathrm{PtSe_{2}}$ than graphene~\cite{Smith2013, Wagner2018} due to the presence of band gap in $\mathrm{PtSe_{2}}$~\cite{Wang2015_2}. \\
\indent In this paper, we intend to enhance the PS of graphene further and bring it at par with that of the TMDs. We explore miniaturization as a means to enhance the performance of graphene membranes for next-generation NEMS pressure sensors. 
The rapid advancement in the state-of-the-art lithography techniques has down-scaled MEMS systems to nanometer range (NEMS)~\cite{Plummer2009}. The membrane theory~\cite{Zhao2008, Timoshenko1959} and thin-film adhesivity model~\cite{Gent1987, Liland2019} predict an increase in the magnitude of strain and critical pressure of the membrane, respectively, with dimension reduction. This will lead to an overall performance enhancement and reduction in the dimension of the graphene pressure sensor. \\
\indent In the subsequent sections, we develop a theoretical model to calculate the pressure-induced strain within a membrane. Using it, we obtain the PS and pressure range (PR) of graphene membranes in the ballistic, quasi-ballistic, and diffusive regimes. The detailed derivations are given in the Appendix sections. 
\section{Theoretical Model} \label{section_2}
\subsection{Simulation setup}
\begin{figure}	
	\subfigure[]{\includegraphics[height=0.14\textwidth,width=0.228\textwidth]{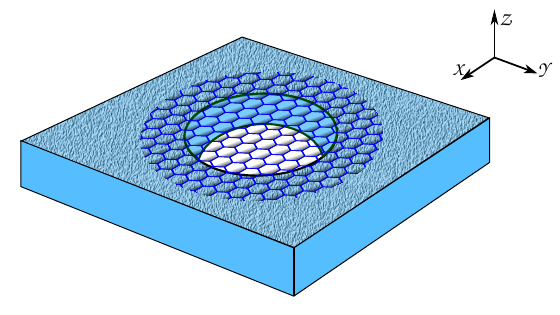}\label{P03_1a}}
	\quad
	\subfigure[]{\includegraphics[height=0.14\textwidth,width=0.228\textwidth]{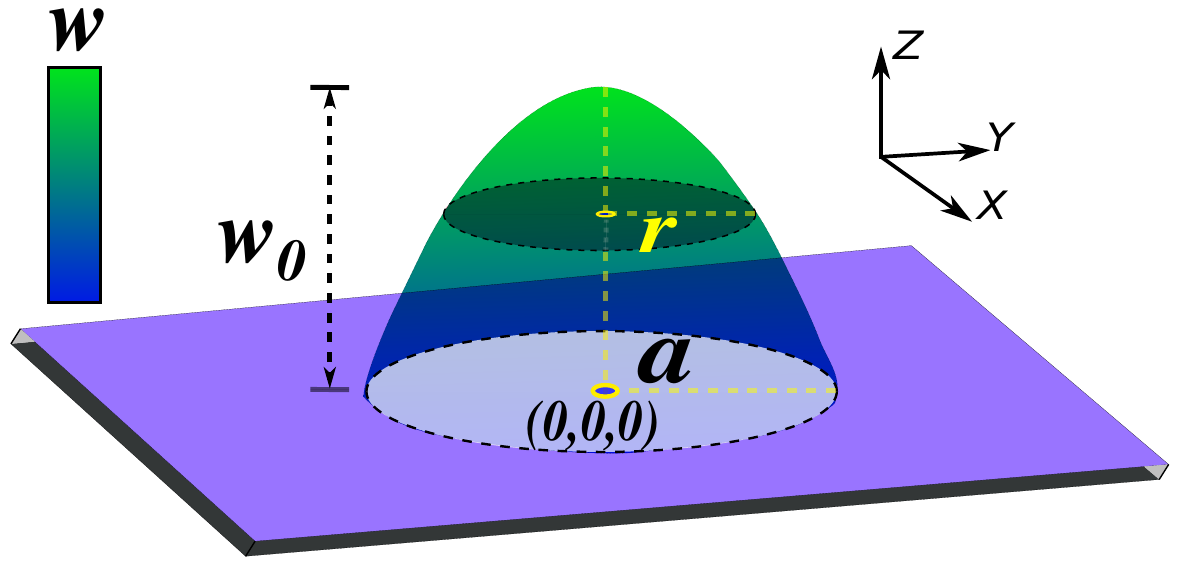}\label{P03_1b}}
	\quad
\caption{(a) Schematic diagram for device setup: a circular graphene membrane of radius `a' on different substrates, used to calculate the pressure sensitivity of graphene. (b) Schematic depicting the deflection of graphene membrane due to an applied pressure across its surfaces.}
\label{P03_1}
\end{figure}

The schematic diagram of the device setup meant to calculate the PS of graphene is depicted in Fig.~\ref{P03_1a}. We use a circular membrane of variable radius $a$ on different substrates to calculate the PS of graphene in different transport regimes, namely ballistic, quasi-ballistic and diffusive regimes. The pressure difference (up to $10^{9}$ Pa) applied across the surfaces of the membrane and the corresponding change in electrical resistance due to the deflection are measured (refer to Fig.~\ref{P03_1b}). The change in resistance with pressure is used to calculate the PS.\\

\subsection{Pressure sensitivity of graphene}

Pressure sensitivity can be expressed as 
\begin{equation}
P.S.=GF\times \frac{\varepsilon}{P},
\label{P03_eq1}
\end{equation}
where $GF$ is the gauge factor, $\varepsilon$ is the strain, and $P$ is the pressure difference across the surface~(refer to Appendix~\ref{P03_app1}). The PS depends on GF and the ratio of strain and pressure. 
We measure the deflection of membrane as a function of pressure using membrane theory~\cite{Timoshenko1959,Zhao2008} and obtain the strain in the membrane. Using the values of strain as a function of pressure, we calculate the PS using the value of GF of graphene in different transport regimes~\cite{Sinha2019, Sinha2020}. Graphene has a very low GF across different transport regimes. Thus, the PS of graphene can be increased by increasing the ratio of $\frac{\varepsilon}{P}$. Since the elastic limit of graphene is very high~\cite{Lee2008}, its adhesivity with the substrate plays a vital role in defining the maximum PR of the graphene sensor.\\
\indent In the subsequent sub-sections, we obtain the mathematical strain as a function of pressure using the deflection of membrane as a function of pressure. Furthermore, we obtain the critical pressure of graphene membranes of different dimensions on various substrates to find the maximum pressure range.

\subsubsection{Deflection of membrane}

The deflection of the membrane ($w$) in the z-direction can be approximated (see Fig.~\ref{P03_1b}) as 
\begin{equation}
w=w_{0}\bigg( 1-\frac{r^2}{a^2}\bigg),
\label{P03_eq2}
\end{equation}
where, $w_{0}$ is the deflection of the center of the membrane, r is the radius of the circle formed in the deflected membrane whose center lies at the coordinates (0,~0,~w), and $a$ is the radius of the blister~\cite{Zhao2008} (see Fig.~\ref{P03_1b}). The value of $w_{0}$ can be calculated by applying the principle of virtual displacement and relevant boundary conditions~\cite{Zhao2008, Timoshenko1959}. Thus, the expression for $w_{0}$ is given by 
\begin{subequations} \label {P03_eq3}
\begin{align}
\frac{w_{0}}{a}=&\bigg(-\frac{\eta}{2} + \zeta \bigg)^\frac{1}{3} + \bigg(-\frac{\eta}{2} - \zeta \bigg)^\frac{1}{3}, \\
\zeta=&\sqrt{\bigg(\frac{\xi}{3}\bigg)^3 + \bigg(\frac{\eta}{2}\bigg)^2},\\
\xi=&\frac{4}{7-\gamma}\bigg(\frac{h}{a}\bigg)^2, and\\
\gamma=& \bigg\{\frac{3(\nu-1)}{7-\nu}\bigg\}\frac{Pa}{Eh},
\end{align}
\end{subequations}
where $\nu$, $h$, $P$, $E$ are respectively the Poisson's ratio, thickness, pressure, and Young's modulus of the graphene membrane.

\subsubsection{Strain along the direction of transport}

\begin{figure}	
\includegraphics[height=0.16\textwidth,width=0.42\textwidth]{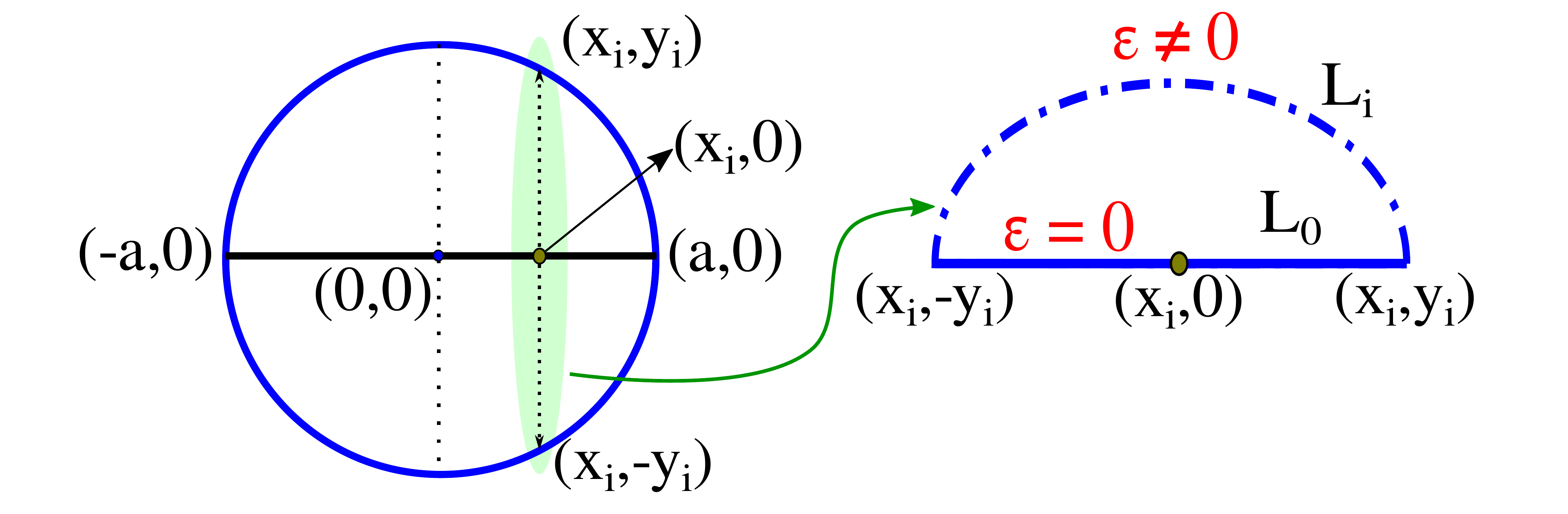}
\caption{Schematic depicting the methodology for average strain calculation in a deflected membrane. The strain value is averaged over $n$ discrete paths. The top and side views of a single path are sketched. }
\label{P03_2}
\end{figure}

\begin{figure}
\subfigure{\includegraphics[height=0.18\textwidth,width=0.225\textwidth]{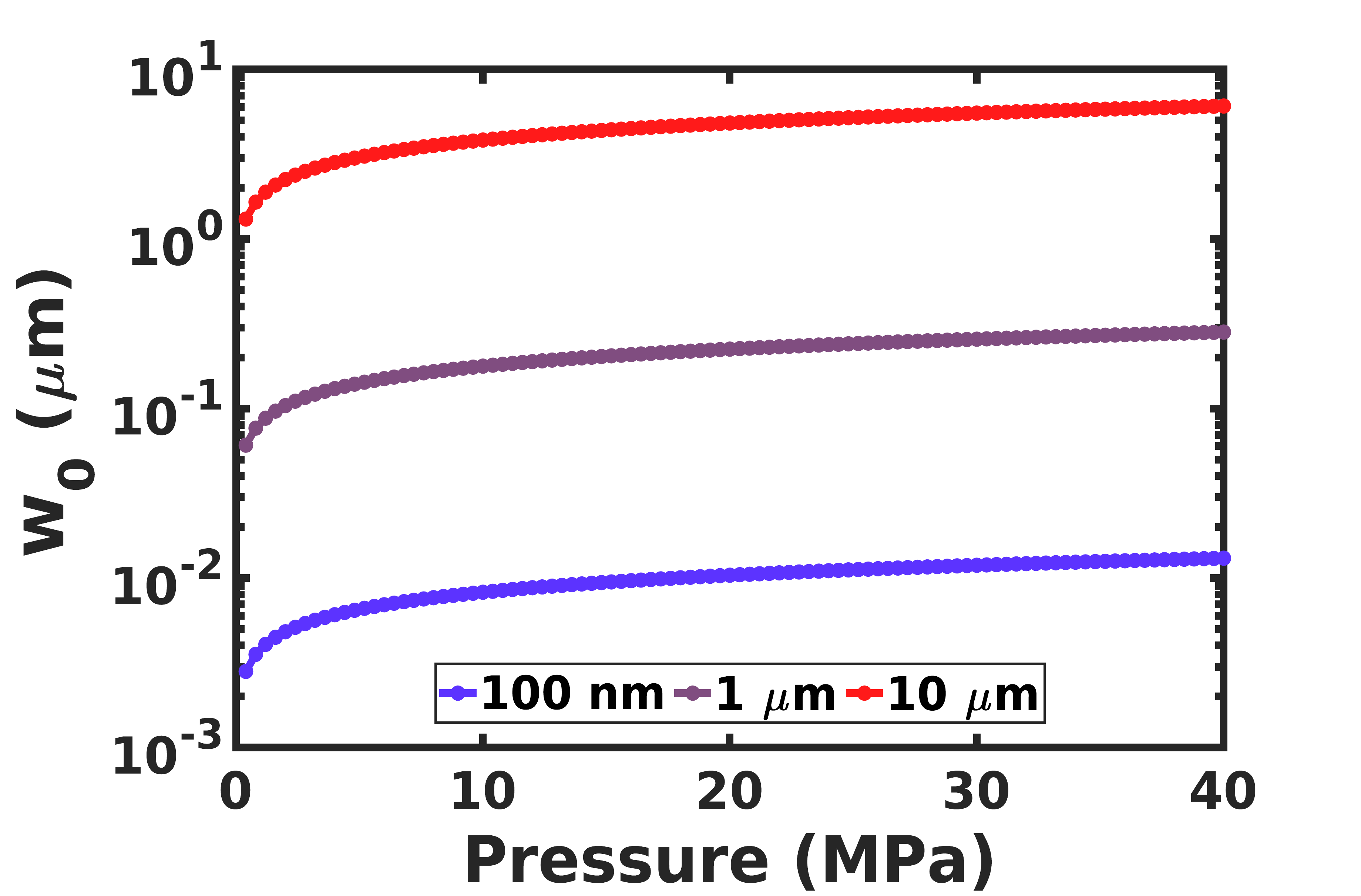}\label{P03_3a}}
\quad
\subfigure{\includegraphics[height=0.18\textwidth,width=0.225\textwidth]{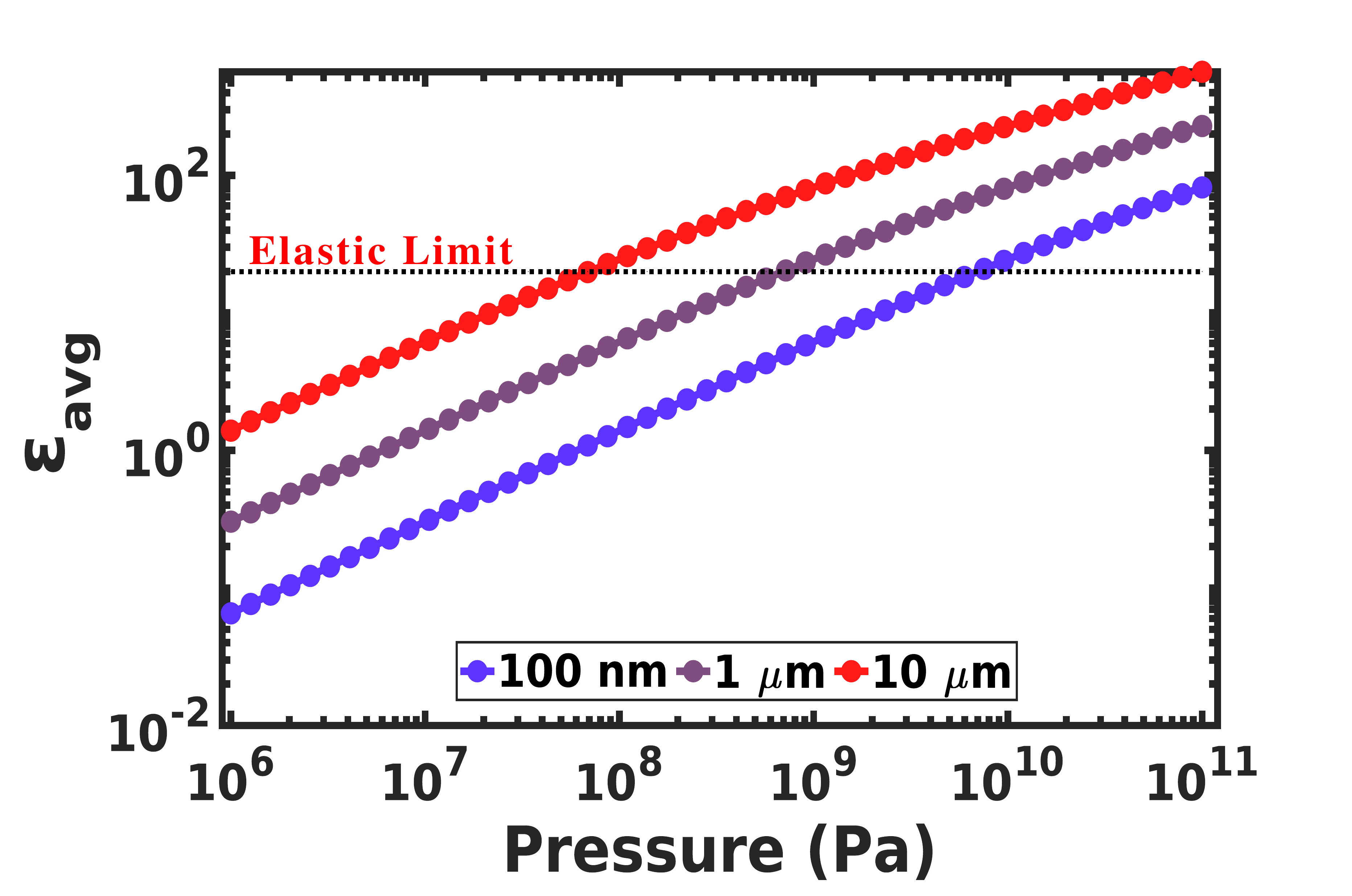}\label{P03_3b}}
\quad
\subfigure{\includegraphics[height=0.18\textwidth,width=0.225\textwidth]{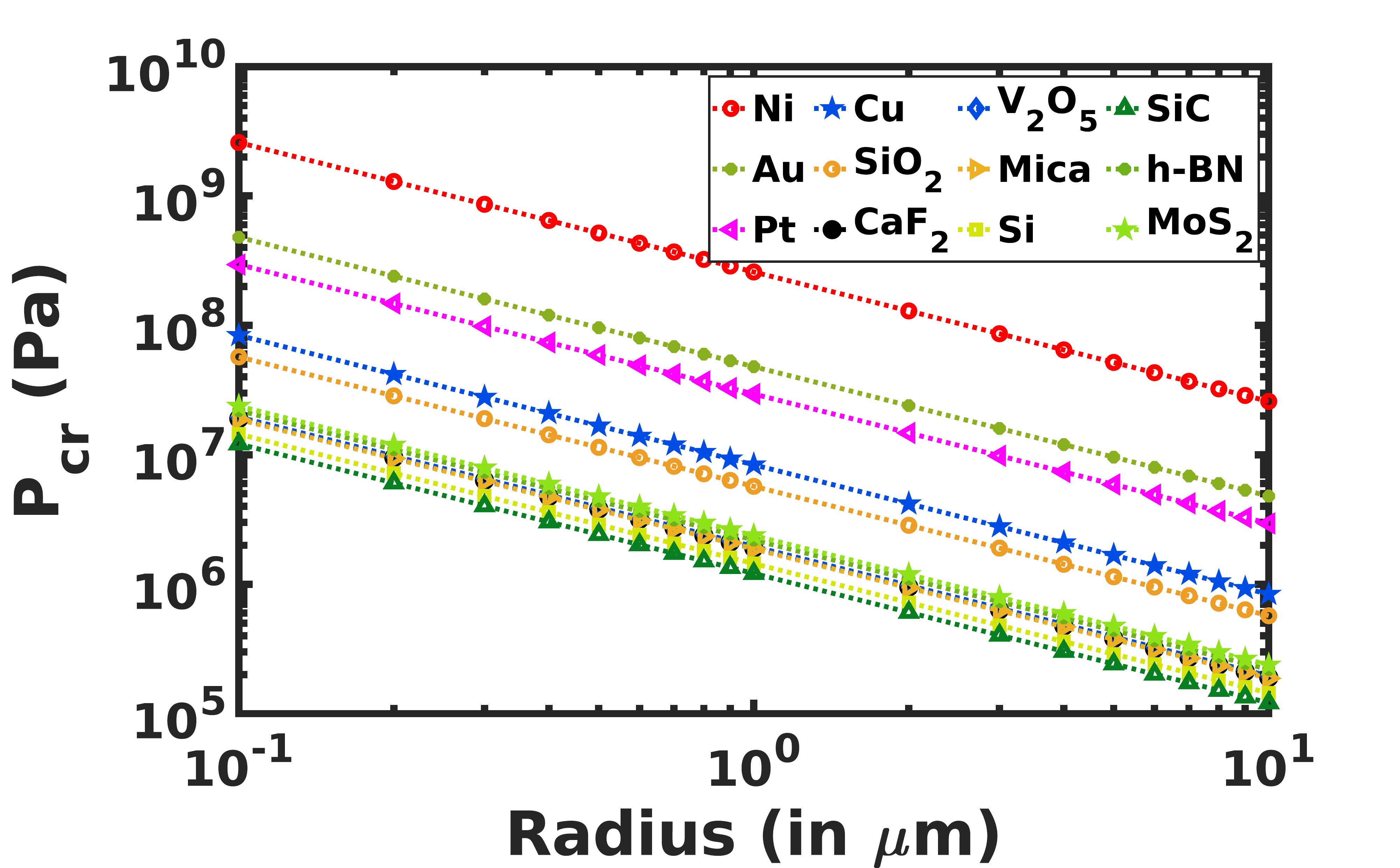}\label{P03_3c}}
\quad
\subfigure{\includegraphics[height=0.18\textwidth,width=0.225\textwidth]{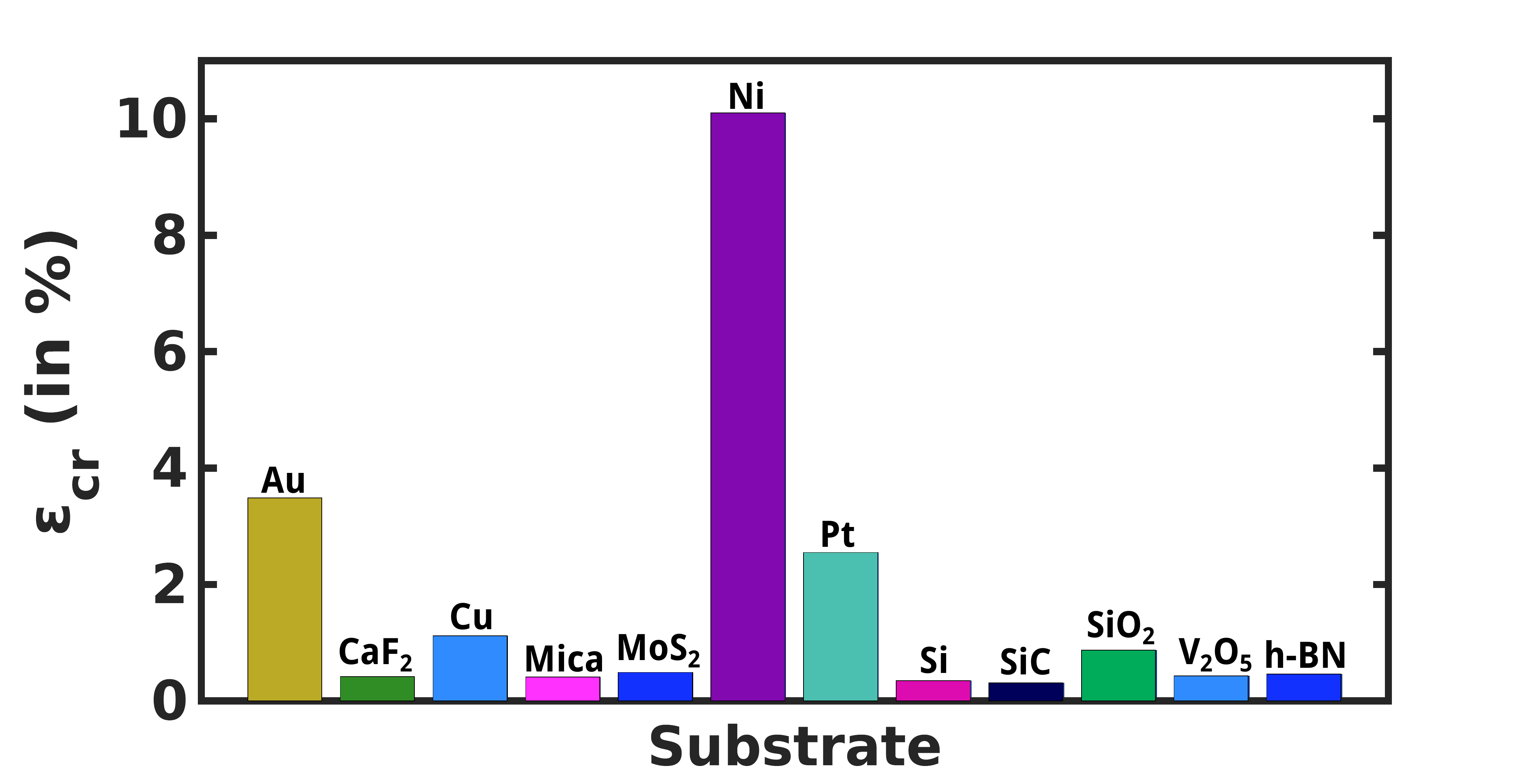}\label{P03_3d}}
\caption{(a) Displacement of the center of membrane ($w_{0}$), and (b) average strain as a function of pressure for different membrane sizes. (c) Critical pressure of graphene sensors as a function of radius for different substrates. (d) Bar plot illustrating the critical strain of graphene on different substrates.}
\end{figure}
An applied pressure across the surface of a circular graphene membrane induces bi-axial strain. In order to calculate PS from uni-axial strain GF, strain along the direction of current flow is required.~\cite{Sinha2019, Sinha2020}. Thus, we obtain a mathematical expression of strain along a specific direction of the deflected membrane. The deflection along various parallel paths, passing through the coordinate $(x_{i},0,0)$ for different values of $x_{i}$ is different (see Fig.~\ref{P03_2}). Thus, it is required to compute the average strain of these paths. The detailed derivation of the expression for average strain along a particular direction is given in Appendix~\ref{P03_app2}. The expression for strain is given by
\begin{subequations} \label {P03_eq4}
 \begin{align}
  \varepsilon_i &=\frac{L_i-L_0}{L_0},\\
  \varepsilon_{i} &= \frac{\sqrt{1+Y^2}}{2} -1 + \frac{1}{2Y}log|Y+\sqrt{1+Y^2}|, \\
  \varepsilon_{avg.} &= \frac{1}{N}\sum\limits_{i=1}^{N} \varepsilon_i, 
 \end{align}
\end{subequations}
where $Y=A y_{i}$, and $A=\frac{-2w_{0}}{a^2}$ and $y_{i}=\sqrt{a^2-{x_{i}}^2}$ (see Appendix~\ref{P03_app3} for detail derivation).
\begin{figure}[]
\subfigure[]{\includegraphics[height=0.18\textwidth,width=0.225\textwidth]{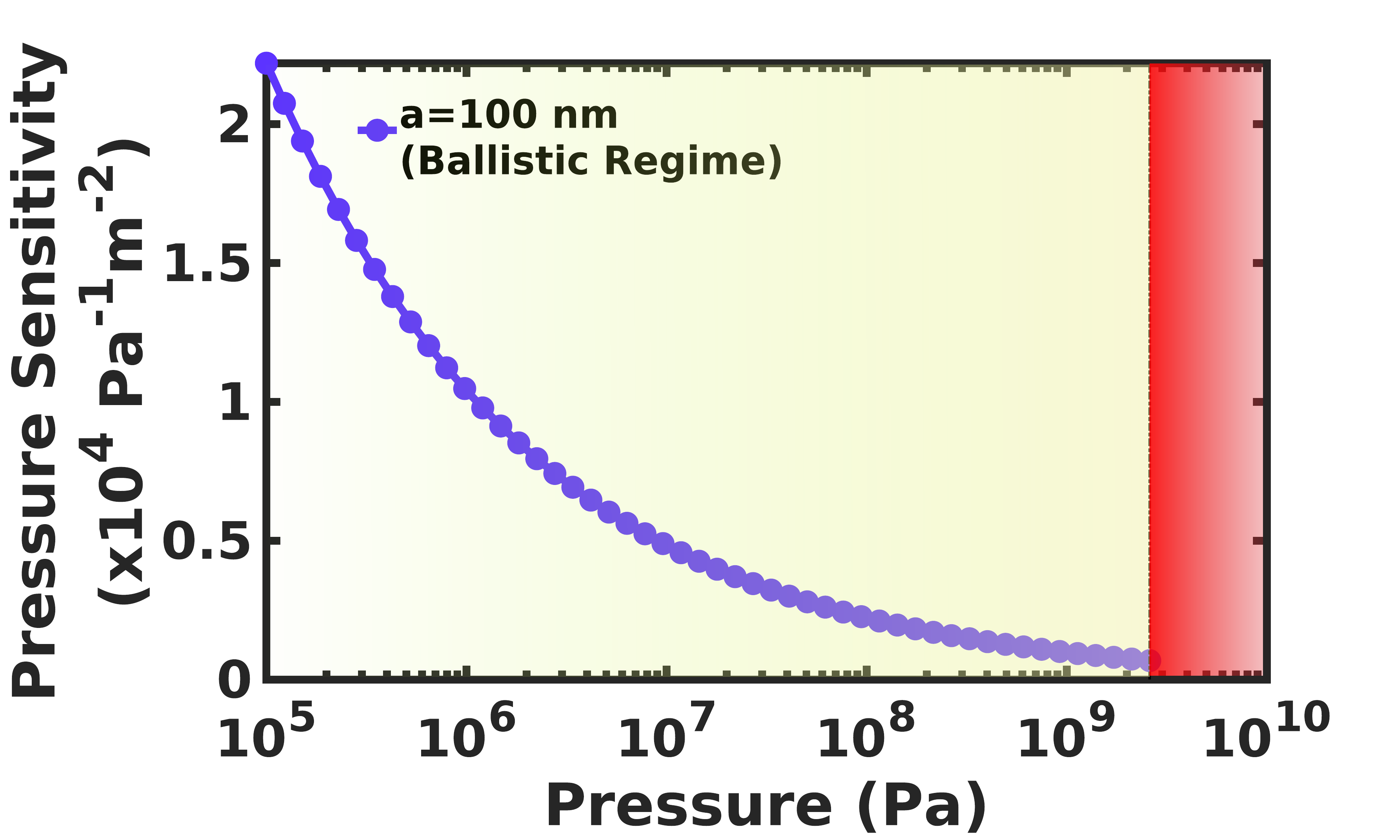}\label{P03_4a}}
\quad
\subfigure[]{\includegraphics[height=0.18\textwidth,width=0.225\textwidth]{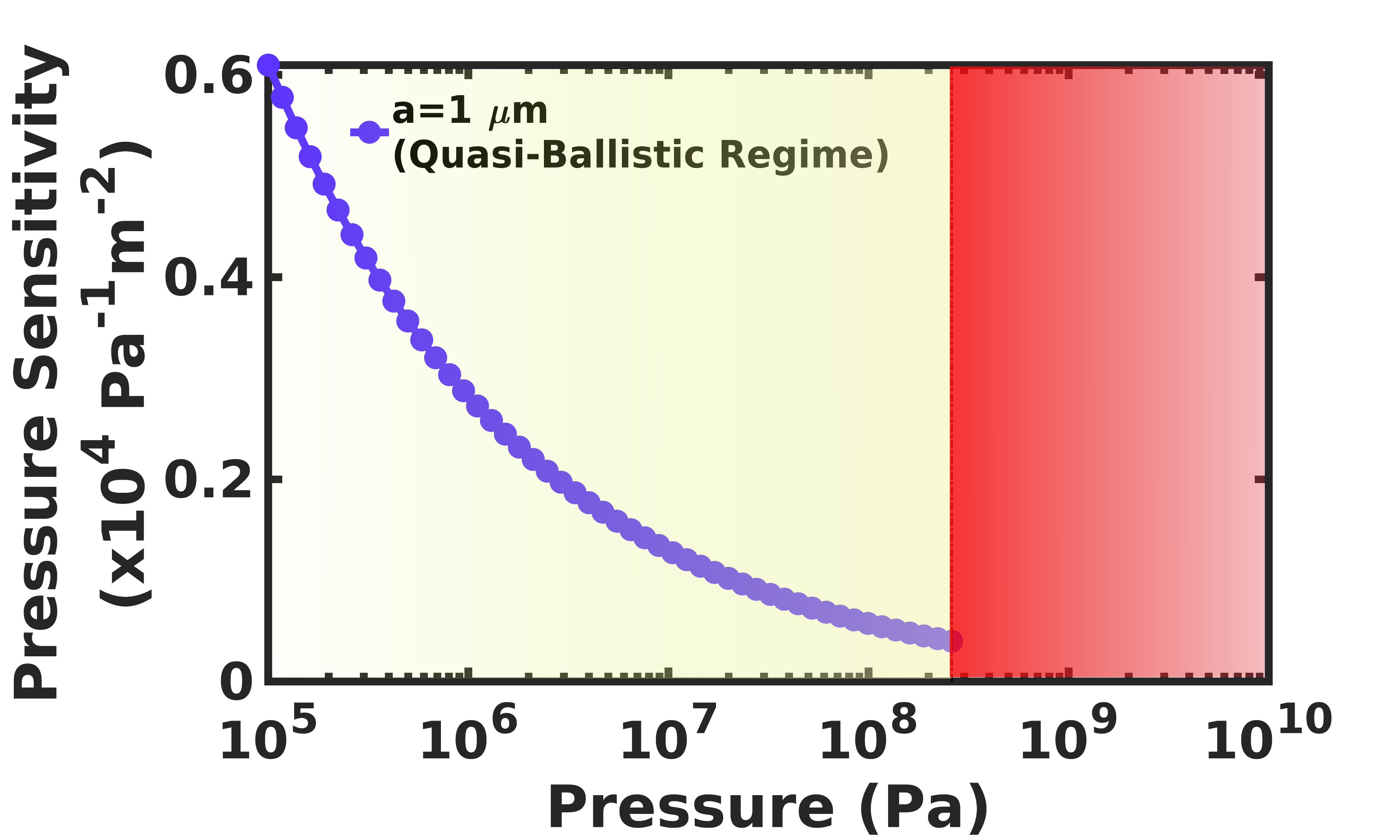}\label{P03_4b}}
\quad
\subfigure[]{\includegraphics[height=0.18\textwidth,width=0.225\textwidth]{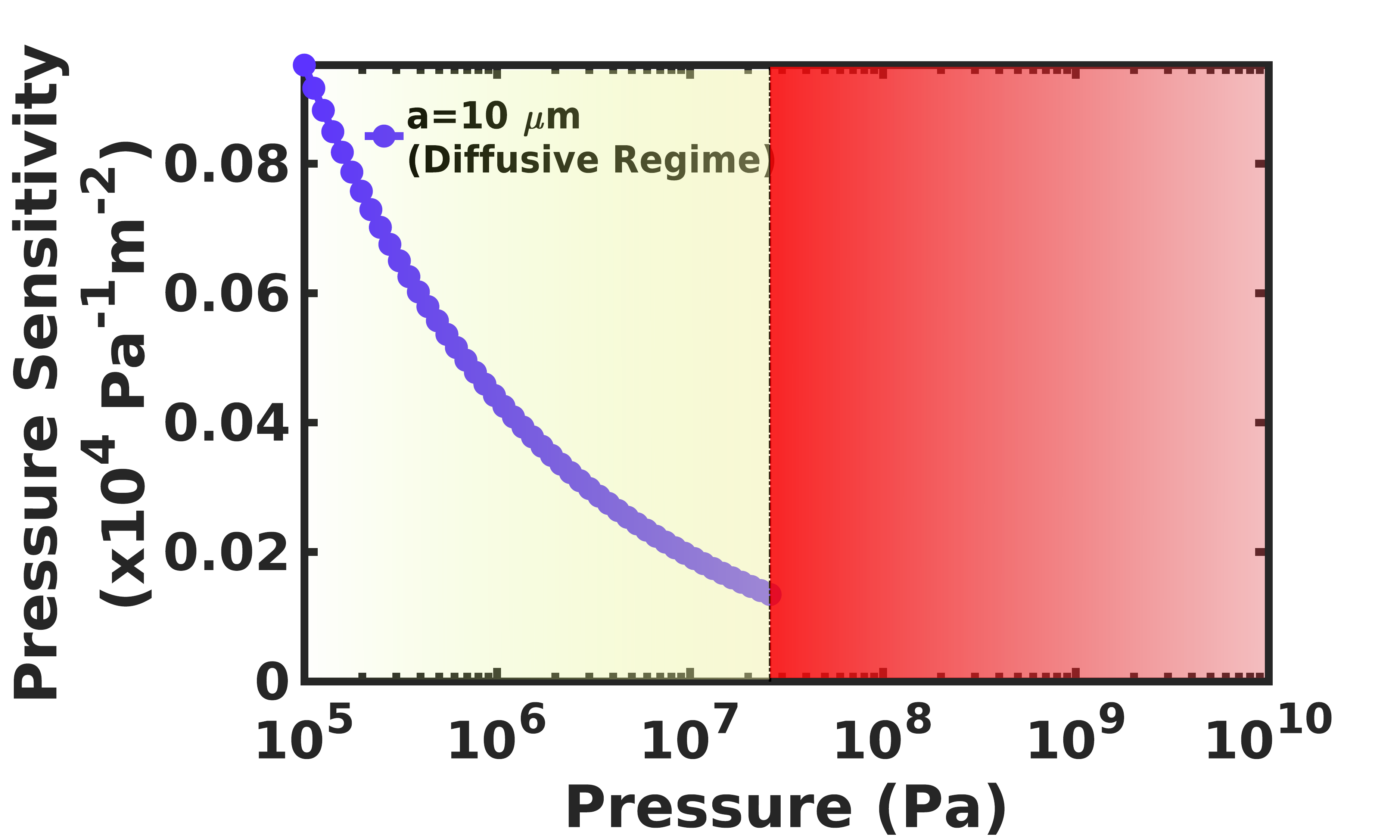}\label{P03_4c}}
\quad
\subfigure[]{\includegraphics[height=0.18\textwidth,width=0.225\textwidth]{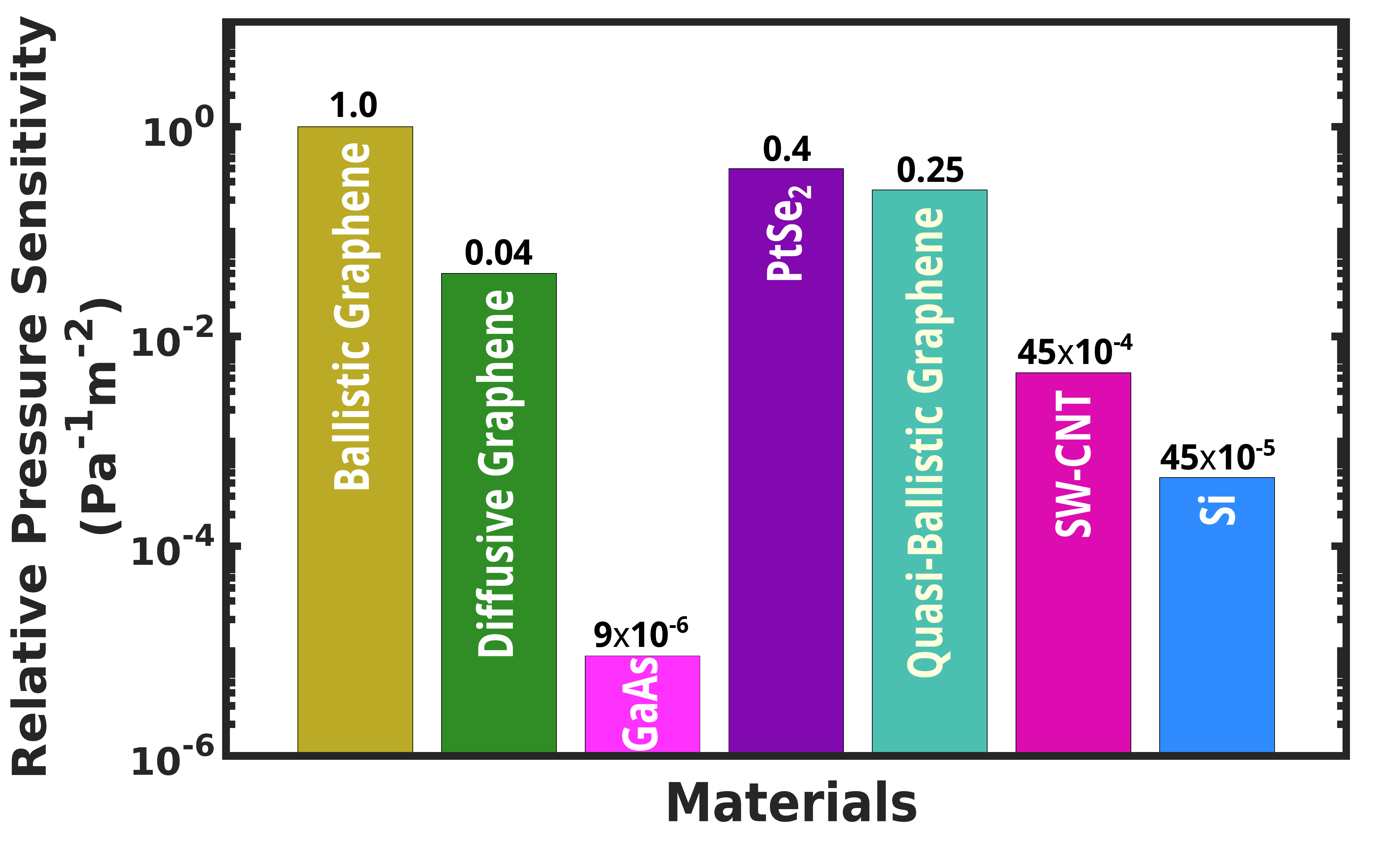}\label{P03_4d}}
\caption{(a)-(c) Pressure sensitivity per unit area as a function of pressure for membrane radii $100~nm$, $1~\mu m$, and $10~\mu m$ respectively. (d) Comparison of normalized pressure sensitivities of ballistic, quasi ballistic, and diffusive graphene sensors with commercially used materials.}
\label{P03_4}
\end{figure}

\begin{table}
	\caption{Comparison between the values of $w_0$ obtained in this work and by Bunch \textit{et. al}~\cite{Bunch2008} at different pressure.}
	\centering
	\begin{tabular}{c@{\hskip 1.1 cm} c@{\hskip 1.1 cm} c} 
		\hline
		\hline						
	$\bm{P~(MPa)}$ & $\bm{w_{0}(nm)}$~(This work) & $\bm{w_{0}(nm)}$~\cite{Bunch2008}  \\	
		\hline			
		 0.145 & 147 & 130 \\
		 0.41 &  207 & 220 \\
		 0.81 &  260 & 280 \\
		\hline
		\hline	
	\end{tabular}	
	\label{P03_table1}
\end{table}

\subsubsection{Adhesivity of graphene}
Graphene is highly adhesive due to its ability to conform to the topography of the substrates~\cite{Koenig2011}. Its high elastic limit~\cite{Lee2008} along with the high adhesivity~\cite{Koenig2011} and impermeability~\cite{Bunch2008} is useful for high-pressure sensing. The maximum PR of graphene depends on its yield-pressure~\cite{Shih2001} as well as critical pressure~\cite{Koenig2011}. \\
\indent The PS can be evaluated using Eq.~\ref{P03_eq1}, once the ratio of average strain and pressure is obtained from Eq.~\ref{P03_eq4}. The critical pressure of a membrane whose thickness is significantly less than its diameter is given by
  \begin {equation}
  P_{cr}= \frac{(17.4EhG^3_{a})^{\frac{1}{4}}}{a},
  \label{P03_eq5}
  \end{equation}
where $P_{cr}$ is the critical pressure, $E$ is Young's modulus, $h$ is the thickness of the membrane, $G_{a}$ is the adhesivity, and $a$ is the radius of the circular membrane~\cite{Gent1987}. Using Eq.~\eqref{P03_eq5}, we obtain the critical pressure of graphene on various substrates. This model accurately predicts the critical pressure of graphene obtained by Koenig \textit{et al.}~\cite{Koenig2011}. The critical pressure obtained experimentally by Koenig \textit{et al.} for a $2.5~\mu m$ radius graphene membrane on $SiO_{2}$ is 1.14 MPa whereas, the critical pressure predicted by Eq.~\eqref{P03_eq5} is 1.4 MPa which is pretty close to the experimentally obtained.  

\section{Results and Discussion } \label{section_3}
\begin{figure}	
\subfigure[]{\includegraphics[height=0.2\textwidth,width=0.225\textwidth]{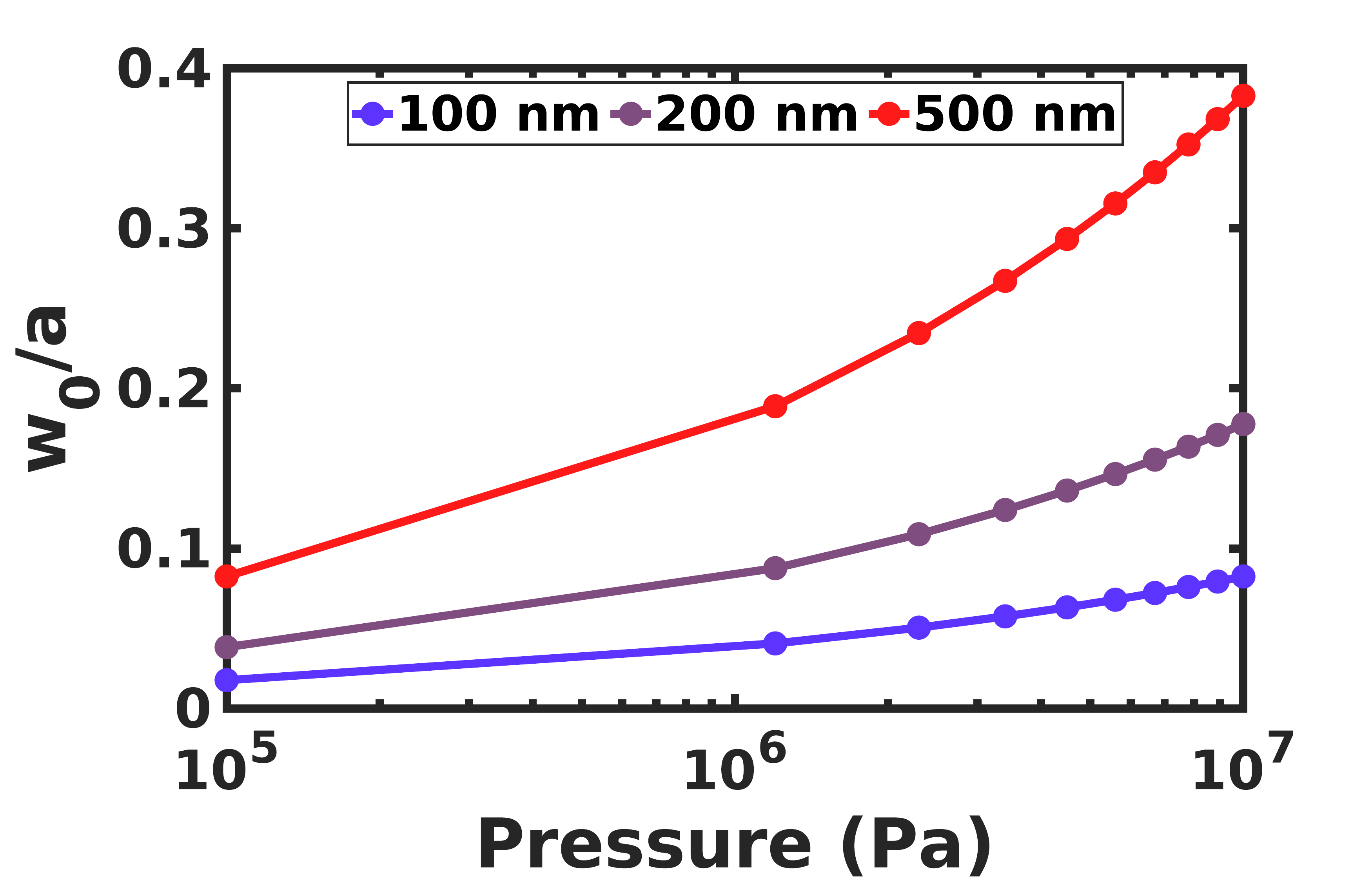}\label{P03_5a}}
\quad
\subfigure[]{\includegraphics[height=0.2\textwidth,width=0.225\textwidth]{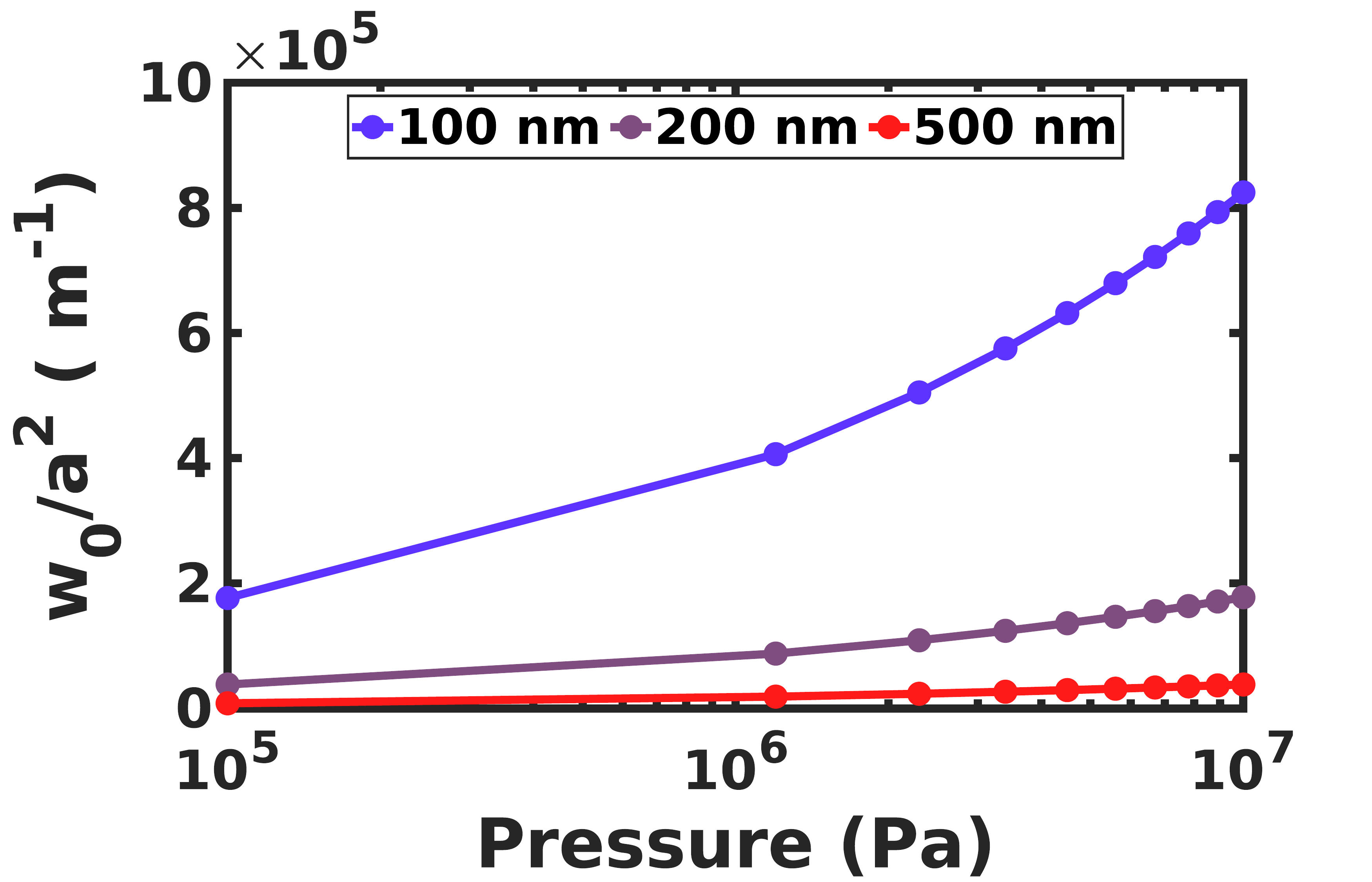}\label{P03_5b}}
\caption{(a) $w_{0}/a$ vs pressure ($P$), and (b) $w_{0}/a^{2}$ vs pressure ($P$) at different values of membrane radius `$a$'. The value of $w_{0}/a$ decreases as the radius `$a$' decreases while, $w_{0}/a^{2}$ increases as the radius `$a$' decreases. Plot (a) implies reduction in PS with decrease in radius `$a$', whereas plot (b) implies an increase in the PS per unit area with decrease in radius `$a$'.}
\end{figure}

\begin{figure*}[]
\subfigure[]{\includegraphics[height=0.25\textwidth,width=0.25\textwidth]{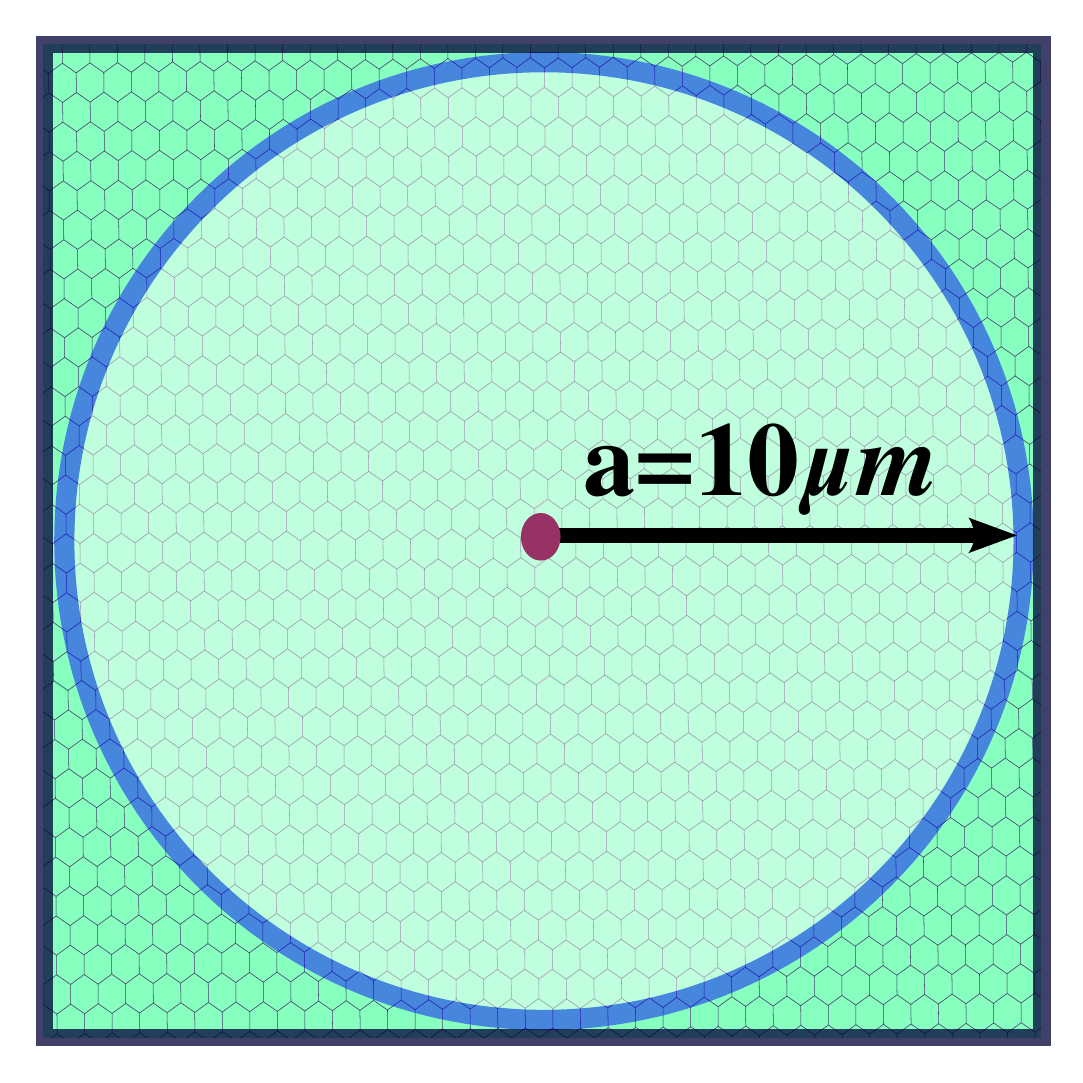}\label{P03_6a}}
\quad
\hspace{0.75 cm}
\subfigure[]{\includegraphics[height=0.25\textwidth,width=0.25\textwidth]{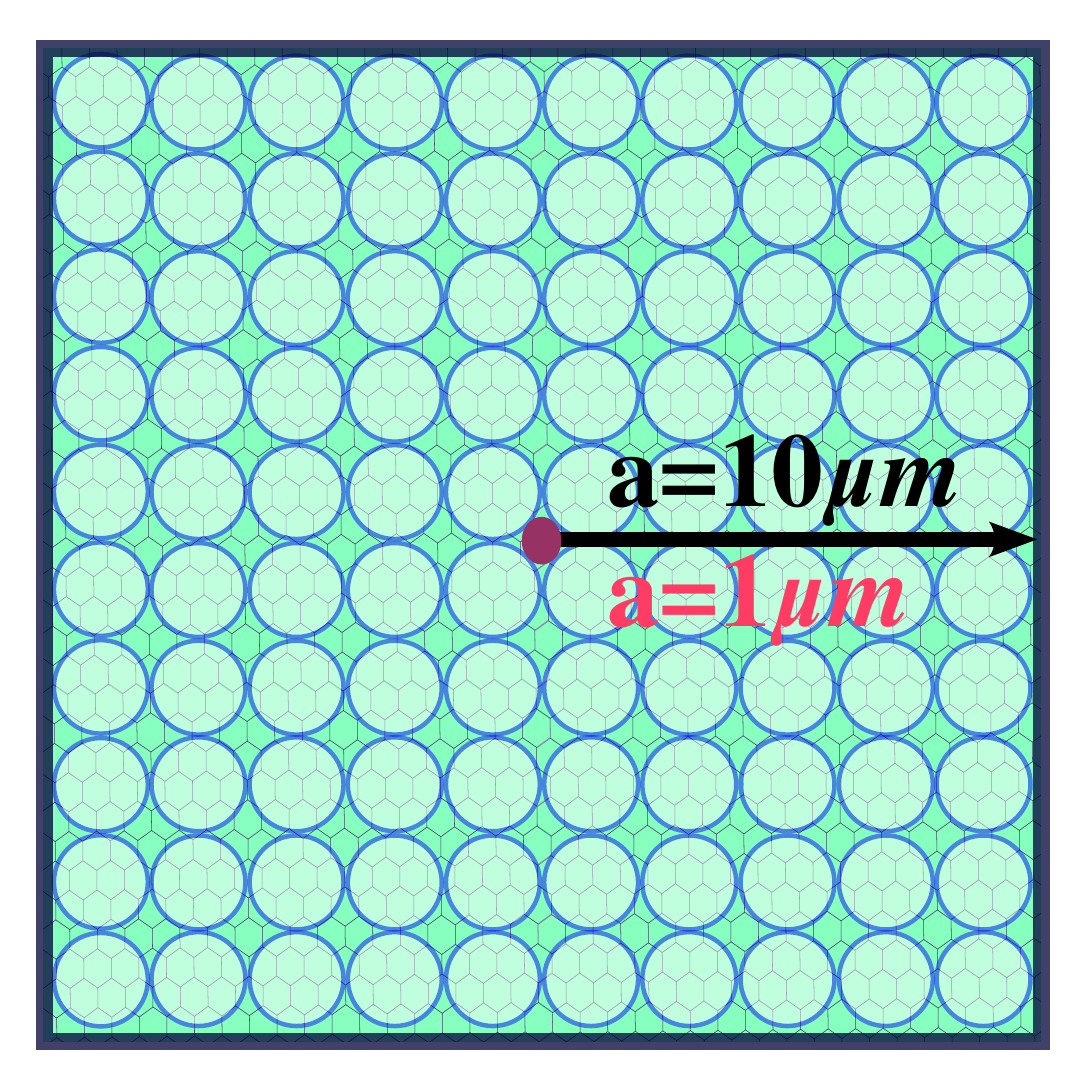}\label{P03_6b}}
\quad
\hspace{0.8 cm}
\subfigure[]{\includegraphics[height=0.25\textwidth,width=0.25\textwidth]{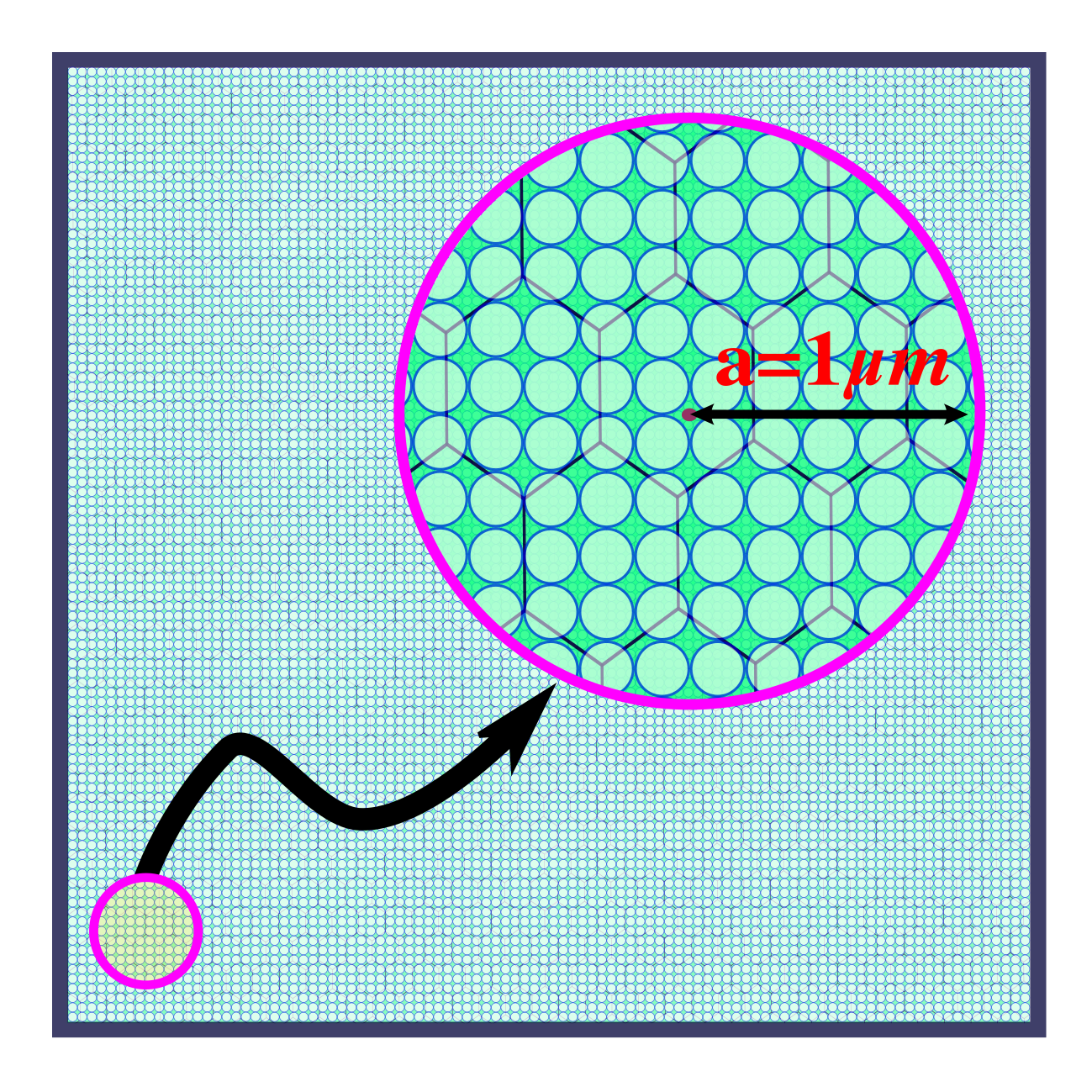}\label{P03_6c}}
\quad
\subfigure[]{\includegraphics[height=0.25\textwidth,width=0.3\textwidth]{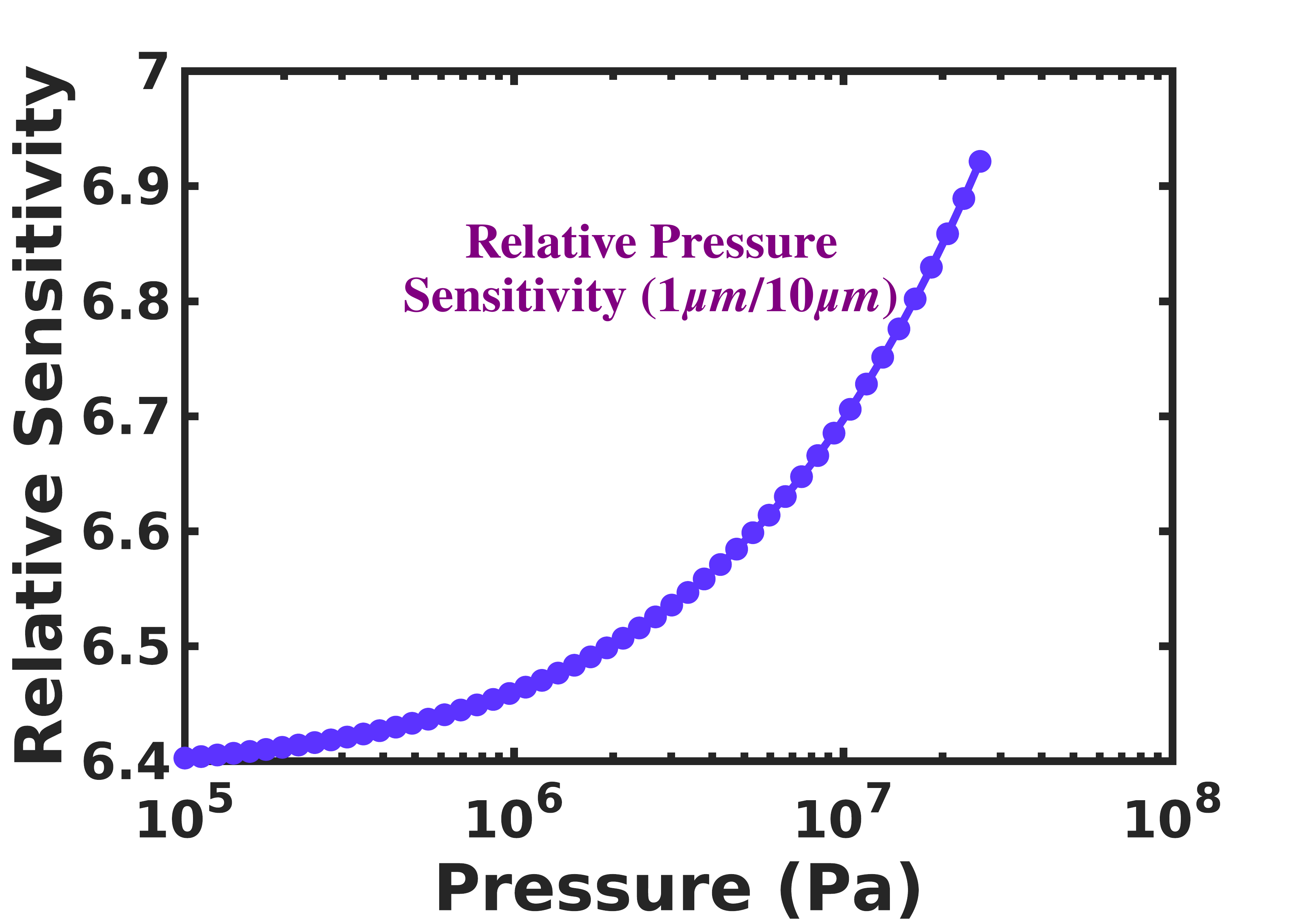}\label{P03_6d}}
\quad
\subfigure[]{\includegraphics[height=0.25\textwidth,width=0.3\textwidth]{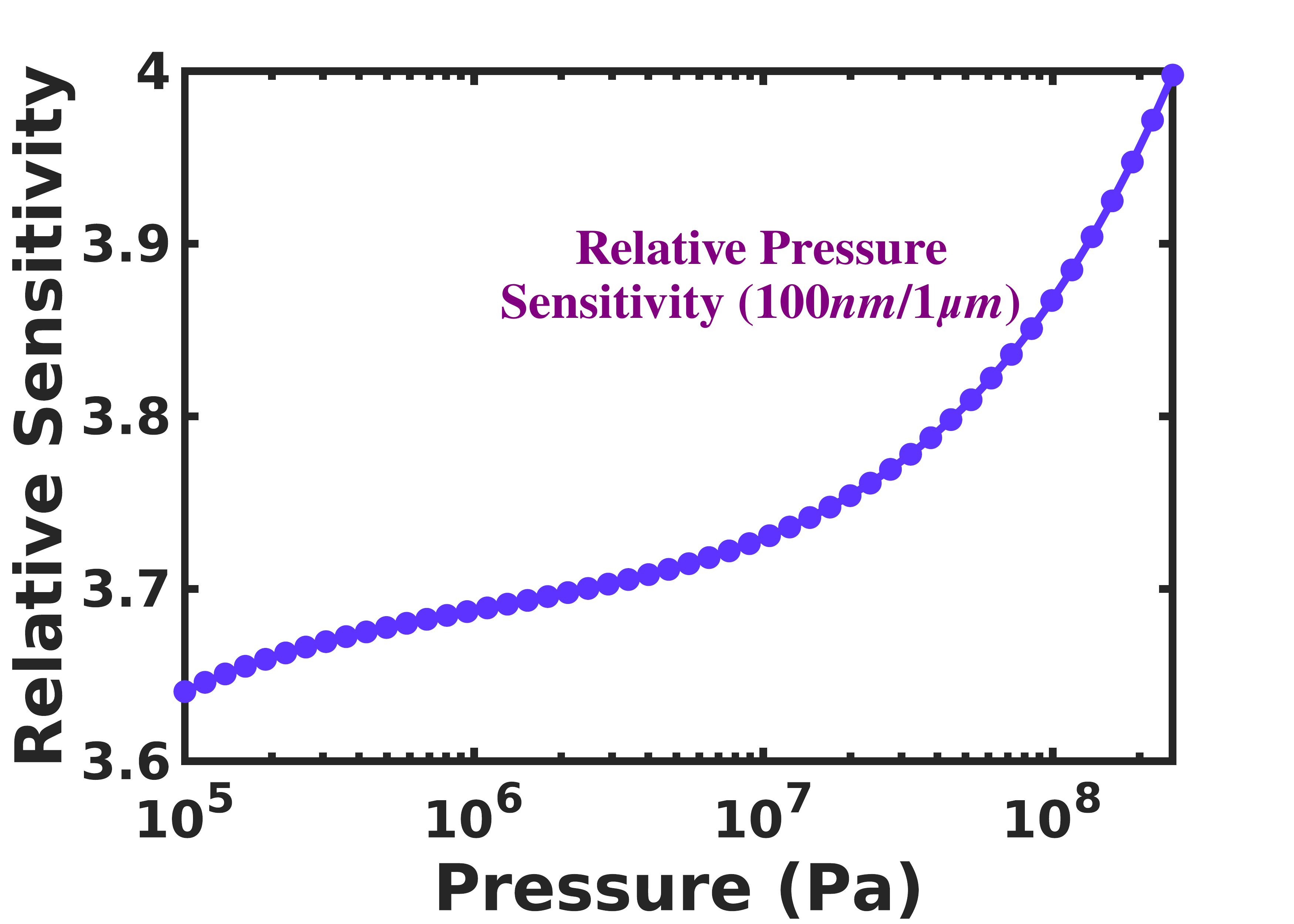}\label{P03_6e}}
\quad
\subfigure[]{\includegraphics[height=0.25\textwidth,width=0.3\textwidth]{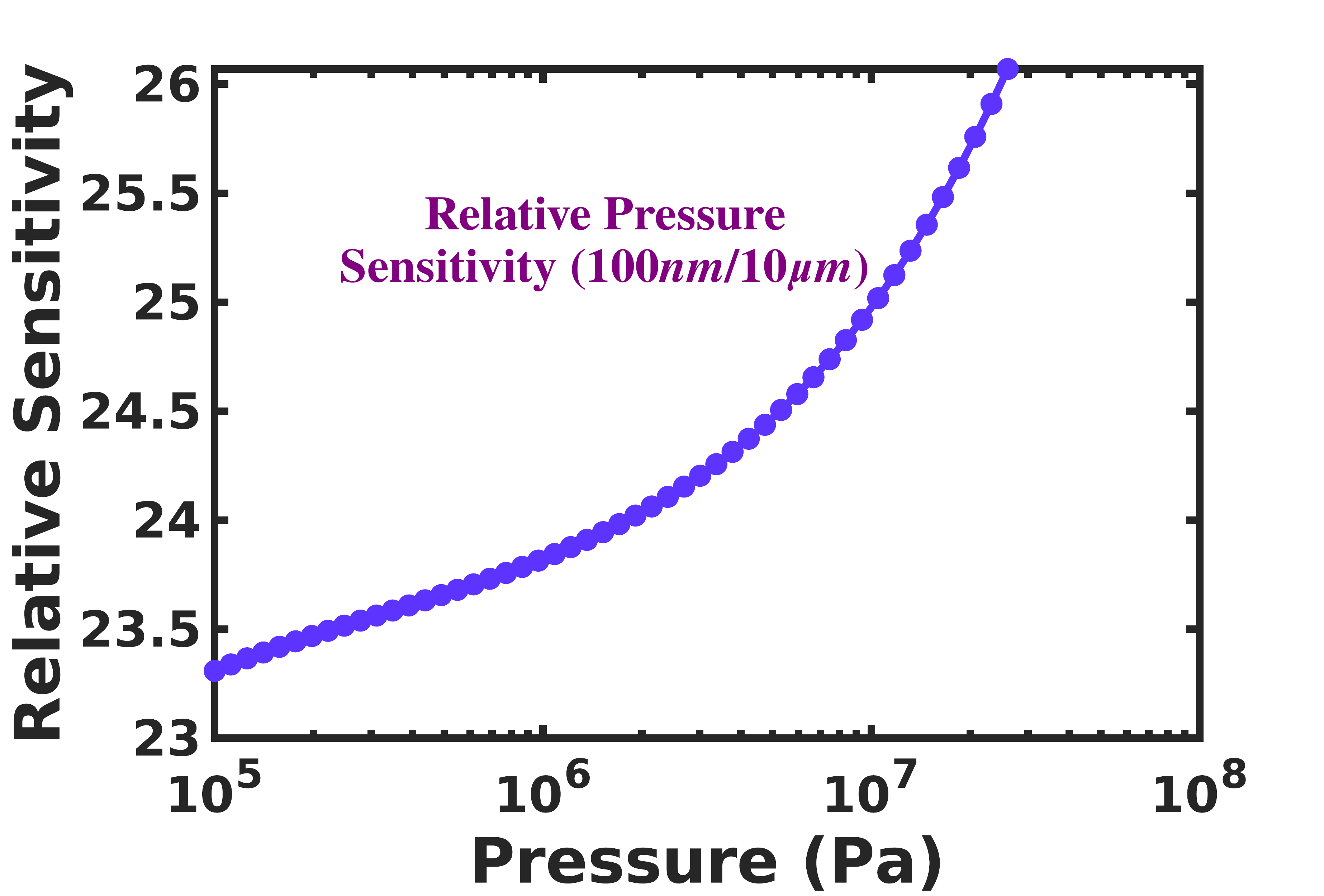}\label{P03_6f}}
\caption{Schematics illustrating (a) a graphene pressure sensor of radius $10 \mu m$ formed from a square sheet of side $20~\mu m$, (b) an array of graphene pressure sensors of radius $100~nm $ or $1~\mu m$ formed from a square sheet of side $2~\mu m$ or $20~\mu m$ respectively, and (c) an array of graphene pressure sensors of radius $100~nm$ formed from a square graphene sheet of side $20~\mu m$. The ratio of pressure sensitivities of: (d) sensors in sub-figures (b) and (a), (e) sensors of sub-figures (b) and (c), and (f) sensors in sub-figures (c) and (a).}
\end{figure*}

\begin{figure*}[ht]	
	\subfigure[]{\includegraphics[height=0.5\textwidth,width=0.7\textwidth]{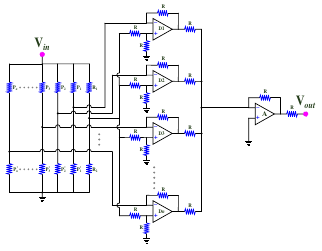}\label{P03_7a}}
	\quad
	\subfigure[]{\includegraphics[height=0.22\textwidth,width=0.3\textwidth]{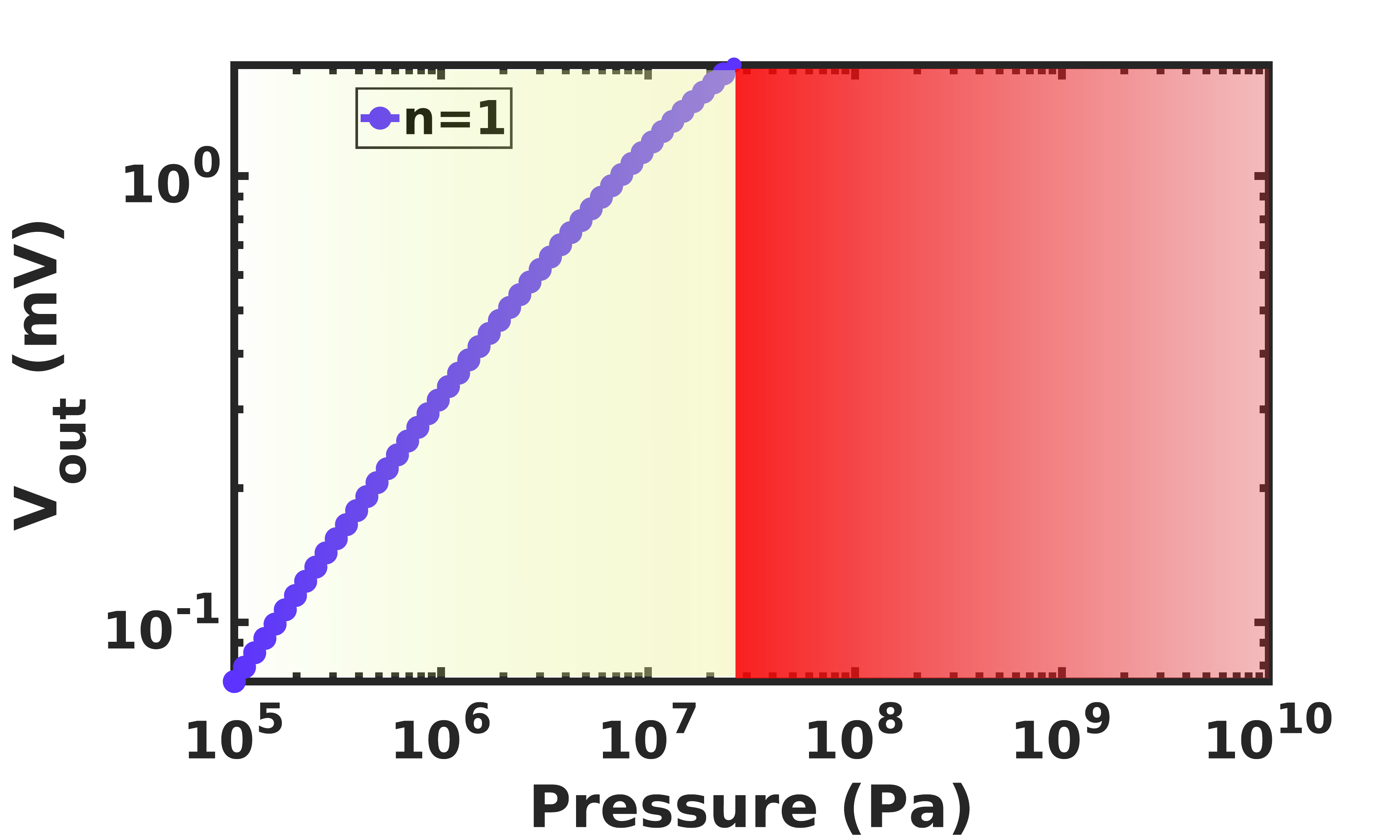}\label{P03_7b}}
	\quad
	\subfigure[]{\includegraphics[height=0.22\textwidth,width=0.3\textwidth]{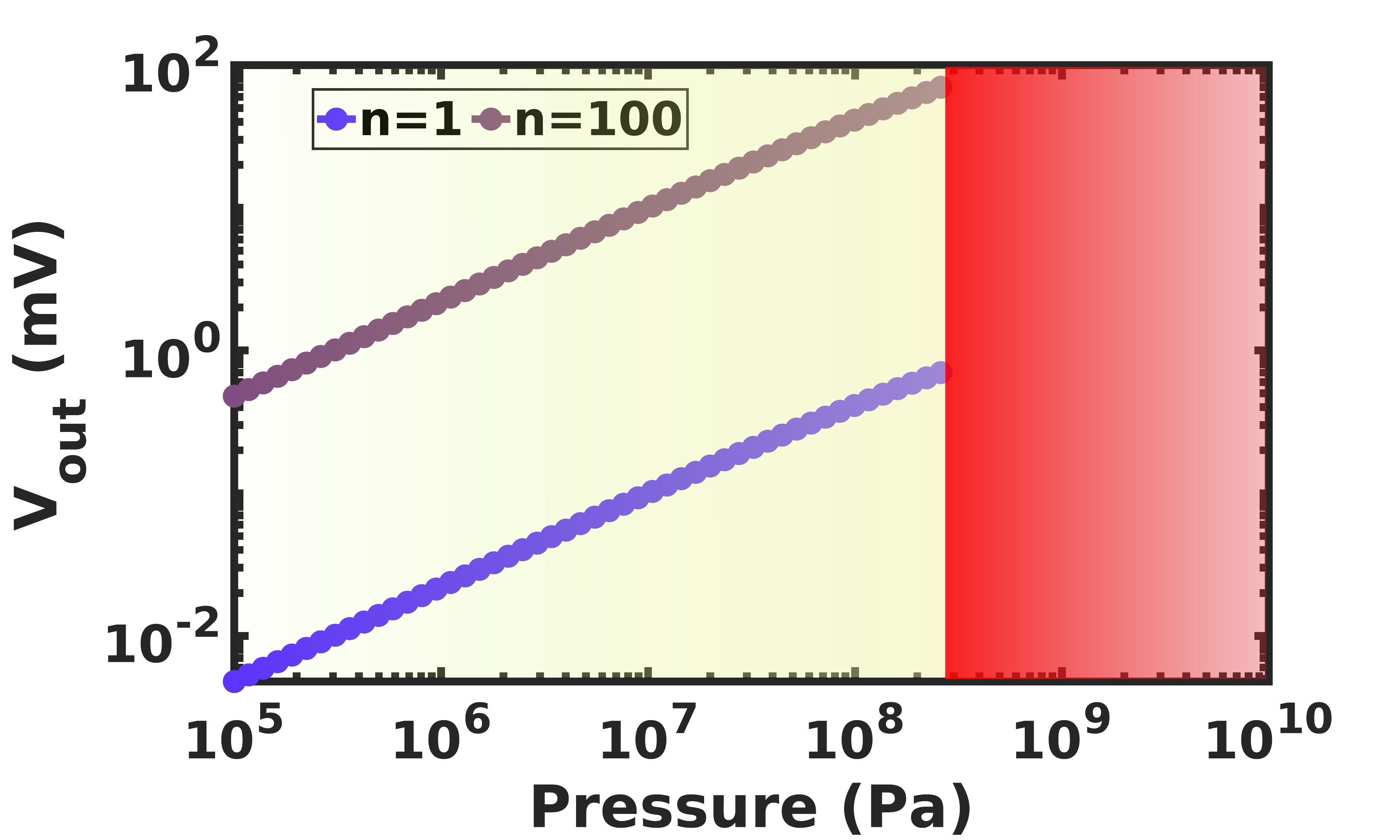}\label{P03_7c}}
	\quad
    \subfigure[]{\includegraphics[height=0.22\textwidth,width=0.3\textwidth]{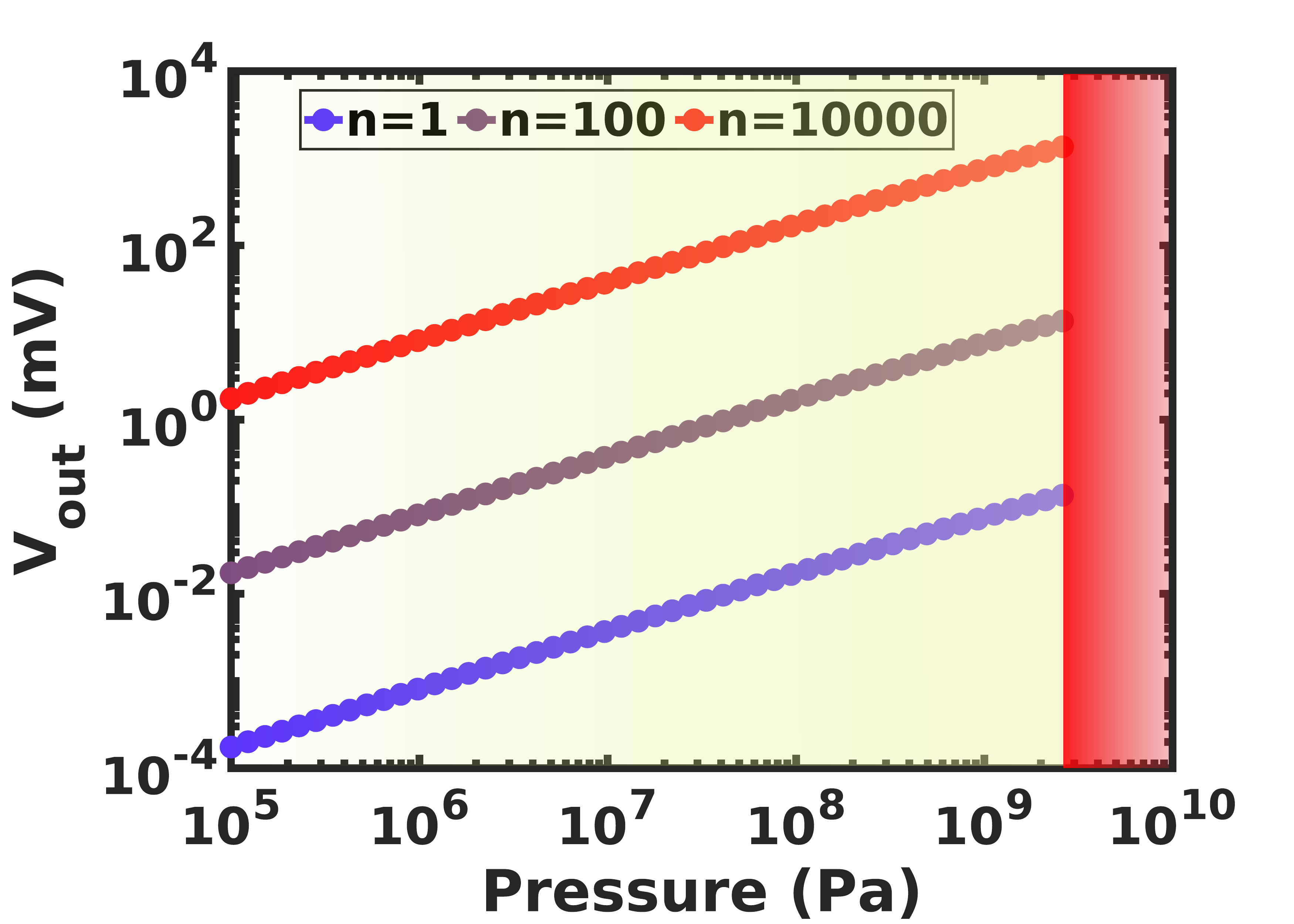}\label{P03_7d}}
	\caption{(a) Circuit diagram of a Multi-channel Wheatstone bridge configuration for graphene array pressure sensors. Output voltage ($V_{out}$) measured using $n$-channel Wheatstone bridge shown in (a) for different number of pressure sensors ($n$) having radius (b) $10 \mu m$, (c) $1 \mu m$ , and (d) $100~nm $ in array configuration. The output voltage and pressure range (represented by yellow shaded region) increase by reducing the size of sensors and increasing the number of sensors in the array, keeping the size of the original graphene sheet intact.}
	\label{P03_6}
\end{figure*}
The performance of a graphene pressure sensor depends on the membrane material and the substrate. The graphene membrane determines the yield pressure (pressure at the yield point), whereas the adhesivity on substrates determines the critical pressure (minimum pressure required to delaminate graphene from the substrate). Because of the high elastic limit (more than $20\%$), yield pressure is much larger than the critical pressure in a graphene pressure sensor. Thus, the PR of a graphene sensor depends on its interaction with the substrate. \\
\indent Figure~\ref{P03_3a} shows an increase in the value of $w_{0}$ with an increase in the membrane radius `$a$', which is evident from Eq.~\eqref{P03_eq3}. Further, in Fig.~\ref{P03_3b}, we observe an increase in the yield pressure with a reduction in the membrane radius. It is because the strain in a deflected membrane is proportional to $\frac{w^{2}_{0}}{a^2}$  at a constant pressure~\cite{Timoshenko1959,Zhao2008} (see \ref{P03_app4}) and the value of $\frac{w^{2}_{0}}{a^2}$ decreases as radius `$a$' decreases (see Fig.~\ref{P03_5a}). Since the value of yield strain of graphene is independent of its dimension. So, the yield point is reached at a higher pressure when its radius is smaller. The yield pressures of $10~\mu m$, $1~\mu m$, and $100~nm$ are of the order of $10^8$~Pa, $10^9$~Pa, and $10^{10}$~Pa, respectively.\\
\indent Further, we obtain the critical pressure of graphene from Eq.~\ref{P03_eq5}, which is inversely proportional to the radius `$a$'. Consequently, we witness an increase in the critical pressure with the reduction in radius (see Fig.~\ref{P03_3c}). Figure~\ref{P03_3c} shows the variation of critical pressure as a function of membrane size and substrate. The figure concludes that critical pressure increases 100 times when its radius reduces from $10~\mu m$ to $100~nm$. Apart from the radius, the critical pressure of graphene strongly depends on the substrate. In general, graphene is more adhesive to metal substrates such as nickel, gold, and copper than amorphous insulators such as $SiO_{2}$~\cite{Megra2019}. Nickel has the highest adhesivity reported so far because of stronger graphene-metal interaction and better lattice match~\cite{Gamo1997}. \\
\indent From Figs.~\ref{P03_3b} and \ref{P03_3c}, we infer that because of the large yield pressure of graphene, the critical pressure determines the upper limit of the PR. The critical strains of graphene on various substrates are shown in Fig.~\ref{P03_3d}. Out of them, graphene has the highest critical strain of  $\approx~10\%$ with nickel which is within the linear elastic limit of graphene~\cite{Pereira2009}. \\
\indent Using Eq.~\ref{P03_eq4}, we obtain the PS of ballistic, quasi-ballistic, and diffusive graphene from the values of $\frac{\varepsilon}{P}$ and GF at different transport regimes. Figure~\ref{P03_4} plots the PS of ballistic, quasi-ballistic, and diffusive graphene and compares their sensitivity with various commercially used membranes. From these plots, we find that ballistic graphene ($a=100~nm$) has a very high PS per unit area of nearly $2.5\times 10^4~Pa^{-1}m^{-2}$. It is approximately twenty-five times more sensitive than diffusive graphene ($10~\mu m$) and five times more sensitive than quasi-ballistic graphene. The miniaturization of graphene increases its PS per unit area and significantly enhances the PR. The PS per unit area is proportional to $\frac{w^{2}_{0}}{a^4}$ (see Appendix~\ref{P03_app4}). The value of $\frac{w^{2}_{0}}{a^4}$ increases as radius `$a$' reduces (see Fig.~\ref{P03_5b}). The maximum pressure that can be sensed using ballistic, quasi-ballistic, and diffusive graphene is of the order of $10^9$~Pa, $10^8$~Pa, and $10^7$~Pa respectively.\\
\indent Despite a very low GF~\cite{Sinha2019, Sinha2020, Smith2013, Smith2016}, graphene has a significantly higher PS per unit area than silicon ($10^{3} - 10^{4}$ times) and GaAs ($10^{5}$ times) pressure sensors due to its high surface-to-volume ratio~\cite{Smith2013}. Hence, other atomically thin 2-D materials are also expected to have a high-PS like graphene or even more~\cite{Wagner2018}. Amongst them, TMDs are considered to be the leading contender for replacing graphene due to their high GF (see Table \ref{P03_table2}). Nevertheless, the mechanical properties of graphene are unparalleled, and no other thin-film material could compete with it so far. The excellent adhesivity~\cite{Bunch2008} and mechanical properties~\cite{Lee2008} of graphene give rise to its high PR. The value of PS per unit area of a 2-D sheet of $\mathrm{PtSe_{2}}$, is $10^4~Pa^{-1}m^{-2}$~~\cite{Wagner2018}. Figure~\ref{P03_4d} shows that the PS of graphene in the sub-micron length scale is comparable to that of the $\mathrm{PtSe_{2}}$, despite the latter having a high GF of -85~\cite{Wagner2018}. Thus, using miniaturization of graphene, we can design pressure sensors that have a very high PR and high PS like the TMDs.\\
\indent The higher value of PS per unit area of ballistic graphene does not guarantee higher PS than quasi-ballistic and diffusive graphene due to its smaller surface area.
We propose a novel way to overcome this limitation by forming a large array of ballistic graphene nano-sensors from a graphene sheet instead of using a single large-sized graphene sensor. By doing so, we effectively increase the overall PS and PR of the graphene sensors. Figures~\ref{P03_4a},~\ref{P03_4b} and~\ref{P03_4c} show the formation of a single pressure sensor of radius $10~\mu m$ from a $20~\mu m$ edge square graphene sheet, formation of an array of pressure sensors of radius $100~nm$ and $1~\mu m$ from a $2~\mu m$ and $ 20~\mu m$ square graphene sheet respectively, and formation of an array of pressure sensors of radius $100~nm$ from a $20~\mu m$ square graphene sheet. The relative PS of array sensors and a single pressure sensor formed from the same graphene sheet is shown in Figs.~\ref{P03_6d},~\ref{P03_6e} and \ref{P03_6f}. The relative PS is highest for $100~nm$ radius pressure sensor formed from a $20~\mu m$ square graphene sheet and is lowest for $100~nm$ radius pressure array formed from a $1~\mu m$ square graphene sheet. Hence, we conclude that the smaller the radius of pressure sensors and the larger the size of the graphene sheet, the more is the PS of the system. The advantage of such a system is that a high-PS and high-PR are simultaneously obtained. Based on these findings, we propose an array sensor of radius $100~nm$ formed from a large graphene sheet connected to a multi-channel Wheatstone bridge. \\
\indent In a $n$-channel Wheatstone bridge, each channel has one pressure sensor (represented by P) and a reference piezoresistor (represented by $\mathrm {P^{'}}$) (see Fig.~\ref{P03_7a}). The combination of $n$ channels is identical to $n$ pressure sensors and reference piezoresistor in series in an array configuration. The resultant differential output signal from each channel of the Wheatstone bridge is added to obtain a larger voltage output ($V_{out}$). The use of array sensors using graphene is highly advantageous. The voltage output of a single graphene pressure sensor of radius $20~\mu m$ is shown in Fig.~\ref{P03_7b}, whereas in Fig.~\ref{P03_7c} array of graphene membranes of radius $1~\mu m$ are shown for $n$=1 and 100 and in Fig.~\ref{P03_7d} array of graphene membranes of radius $100~nm $ are shown for $n$=1, 100 and 10000. These figures conclude that the output voltage of smaller-sized array sensors made from a larger sheet when used in the multi-channel Wheatstone bridge configuration is many folds more sensitive than a single pressure sensor made from an identical graphene sheet. We notice a sharp increase in its PR (yellow shaded region) also when multiple smaller sensors are used instead of a single large graphene sensor is used. \\
\indent Piezoresistive pressure sensor transforms pressure waves (sound waves are longitudinal pressure waves) into electrical signals, which is very similar to how a microphone works. Since graphene microphones have the advantage of being resonance-free in the acoustic range~\cite{Wittmann2019}, thus, ballistic graphene array nano-sensors can be instrumental in designing highly sensitive microphones with high pitch and can eventually be useful in modern smart phones.\\
\indent Apart from this, the methodology used for analyzing ballistic graphene pressure sensors in this paper can be used to explore similar applications in other 2-D materials. Amongst 2-D materials, Dirac materials have lower GF than TMDs due to the robust Dirac cones~\cite{Hosseini2015}. Hence, it is more likely that nanoscale TMD sheets will have a larger PS than graphene, despite a smaller PR due to lower elastic limit of TMDs than graphene (see Table~\ref{P03_table2}). 
\begin{table*}
\caption{Thickness, Poisson's ratio, Young's modulus, Elastic limit, and Gauge factor of some common 2-D materials}
\centering
\begin{tabular}{c@{\hskip 0.2 cm} c@{\hskip 0.5 cm} c@{\hskip 0.6 cm} c@{\hskip 0.5 cm} c@{\hskip 0.5 cm} c@{\hskip 0.2 cm}} 
\hline
\hline						
$\bm{Materials}$ & $\bm{h~(nm)}$ & $\bm{\nu} $ & $\bm{E~(N/m)}$ & $\bm{\varepsilon~(\%)}$ & $\bm{GF}$ \\		
\hline	
$\bm{Graphene}$ & 0.335~\cite{Novoselov2004,CastroNeto2009} & 0.14~\cite{Ribeiro2009}  & 335~\cite{John2016}  & $>20$~\cite{Lee2008} & 0.3-10.3~\cite{Sinha2019,Sinha2020,Smith2013}   \\
$\bm{Silicene}$ & 0.313~\cite{Peng2013_2} & 0.31~\cite{Qin2014}  & 61.33~\cite{John2016}  & 5~\cite{Peng2013_2} & -1.4 \\
$\bm{Phosphorene}$ & 0.9~\cite{Liu2014_3}& 0.70(zz),0.18(ac)~\cite{Wang2015_1} & 92 (zz),23 (ac)~\cite{Wei2014,Wang2015_1} & 2~\cite{Liu2019} & 120~\cite{Zhang2017,Nourbakhsh2018}\\
$\bm{MoS_2}$ & 0.65~\cite{Li2015} & 0.25~\cite{Kang2015} & 175.5~\cite{Bertolazzi2011}  & 3~\cite{Peng2013} & -148~\cite{Manzeli2015}  \\
$\bm{MoSe_2}$ & 0.8~\cite{Lu2014} & 0.25~\cite{Yoo2018}  & 80~\cite{Wang2019}  & 2.5~\cite{Yang2017}  & 1700~\cite{Hosseini2015}\\
$\bm{WS_2}$ & 0.9~\cite{Park2017} & 0.23~\cite{Yoo2018} & 170~\cite{Liu2014_2} &2~\cite{Deng2018_2} & 70~\cite{Zhang2021} \\
$\bm{WSe_2}$ & 1.5~\cite{Chen2018} & 0.21~\cite{Yoo2018} & 132.8~\cite{Zeng2015} & 2.5~\cite{Ding2019} & 3000~\cite{Hosseini2015} \\
$\bm{PtSe_2}$ & 0.507~\cite{Kliche1985} & 0.25~\cite{Du2018} & 58.76~\cite{Deng2018_1} & 2~\cite{Du2018} & -85~\cite{Wagner2018} \\
$\bm{ReSe_2}$ & 0.66~\cite{Yang2014} & 0.223~\cite{Wang2017} & 92.4~\cite{Wang2017} & 3~\cite{An2019} & 50~\cite{An2019},-60~\cite{An2019} \\
\hline
\hline	
\end{tabular}	
\label{P03_table2}
\end{table*}
\section{Conclusion} \label{section_4}

Motivated by the recent findings that showed the high-pressure sensitivity of atomically thin materials such as graphene, we explored the pressure sensitivity of miniaturized graphene membranes on various substrates using the membrane theory and thin-film adhesivity model. We used these findings on graphene membranes of different dimensions on various substrates and found a $10^{3}$ fold enhancement in the normalized pressure sensitivity per unit area of ballistic graphene compared to commercial silicon pressure sensors. Apart from pressure sensitivity, we showed that ballistic graphene could sense ultra-high pressure of the order of $10^{9}$ Pa. Moreover, we also found that ballistic graphene is 2-3 times more sensitive than TMD, such as $\mathrm{PtSe_{2}}$, despite the latter having a much higher gauge factor compared to graphene. Finally, this paper concludes that a ballistic graphene pressure sensor is a perfect blend of high-pressure sensitivity and ultra-high pressure range. Based on these results, we proposed ballistic graphene array sensors as next-generation NEMS pressure sensors for ultra-high pressure sensing.

\begin{acknowledgments}

The Research and Development work undertaken in the project under the Visvesvaraya Ph.D. Scheme of Ministry of Electronics and Information Technology, Government of India, is implemented by Digital India Corporation (formerly Media Lab Asia). This work was also supported by the Science and Engineering Research Board (SERB), Government of India, Grant No. CRG/2021/003102 and Grant No. STR/2019/000030.

\end{acknowledgments}
\appendix 

\section{{Expression for pressure sensitivity} \label{P03_app1}}

Pressure sensitivity is expressed as 

\begin{equation}
    PS=\frac{\Delta R/R}{P},
    \label{P03_app1_eq1}
\end{equation}
where $\Delta R$ is the change in resistance due to pressure $P$ and $R_{0}$ is the resistance of membrane at zero strain. 
The right hand side of Eq.~\ref{P03_app1_eq1} can be written as $\frac{\Delta R}{\varepsilon R}\times \frac{\varepsilon}{P}$. The quantitiy $\varepsilon$ is the strain generated in the membrane due to pressure ($P$) and $\frac{\Delta R}{\varepsilon R}$ is the gauge factor (GF). Thus, Eq.~\eqref{P03_app1_eq1} simplifies into 
\begin{equation}
    PS= GF \times \frac{\varepsilon}{P}
\end{equation}

\section{{Deflection of membrane}\label{P03_app2}}
The pressure difference across the top and bottom surfaces causes the membrane to deflect (see Fig.~\ref{P03_1b}). The large scale deflection of a membrane is approximated by the expression 
\begin{equation}
 w=w_{0}\bigg( 1-\frac{r^2}{a^2}\bigg).
\end{equation}
The radial displacement (u) can be approximately written as 
\begin{equation}
    u=r(a-r)(c_1 + rc_2),
\end{equation}
where $c_1$ and $c_2$ are constant.\\
The value of $w_0$ can be easily computed using the principle of virtual displacement. According to this principle, change in mechanical energy is equal to work done by the pressure to deflect the membrane~\cite{Timoshenko1959}. The mechanical energy of a membrane includes strain energy and bending energy. On solving the value of $w_0$ using this principle, we obtain an analytical expression of $w_{0}$ which is a function of  applied pressure, thickness, and radius of the membrane~\cite{Zhao2008,Timoshenko1959}.\\
\indent The value of strain energy of a circular membrane is given by
\begin{subequations} 
 \begin{align}
    V_{s}  = \pi K\int\limits_0^a \big( \varepsilon_r^2 + & \varepsilon_{\theta}^2 +2\nu \varepsilon_r  + \varepsilon_{\theta} \big)rdr, \\
    K =& \frac{Eh}{1-\nu^2},\\
    \varepsilon_r = & \frac{du}{dr} + \frac{1}{2}\bigg(\frac{dw}{dr}\bigg)^2, and \\
        \varepsilon_{\theta}= & \frac{u}{r}.
    \end{align}
\end{subequations}
On simplifying the above equations, we obtain the strain energy as
\begin{equation}
    \begin{split}
    V_{s}= &\frac{\pi Eh}{1-{\nu}^2}\Bigg[ \frac{1}{4}c_1^2a^4 +  \frac{3}{10}c_1c_2a^5 -  \frac{1}{5}(3-\nu)c_1aw_0^2\\ 
      &  + \frac{7}{60}c_2^2a^6 -\frac{2}{15}(3-\nu)c_{2}a^2w_0^2+\frac{2}{3}\frac{w_0^4}{a^2}       \Bigg]
    \end{split}
\end{equation}
The bending strain energy of a membrane is given by~\cite{Timoshenko1959}
\begin{equation}
    V_{b}=4\pi D (1+\nu) \bigg( \frac{w_0}{a}\bigg)^2,
\end{equation}
where $D=\frac{Eh^{3}}{12(1-\nu^{2})}$ is the bending rigidity. The work done by the pressure while deflecting the circular membrane is given by~\cite{Zhao2008}
\begin{equation}
W=\frac{\pi a^2 P}{2} w_{0}
\end{equation}

In equilibrium, the value of $V_s$ is minimum with respect to $c_1$ and $c_2$. Using these conditions along with the principle of virtual displacement~\cite{Timoshenko1959}, we obtain the value of $w_0$. The conditions are 
\begin{subequations} 
    \begin{align}
    \frac{dV_s}{dc_1} = \frac{dV_s}{dc_2} & = 0, \\
    \frac{d(V_{s} + V_{b})}{dw_{0}}\delta w_{0} & =  \frac{dW}{dw_{0}}\delta w_{0}.
   \end{align}
   \label{P03_eq7}
\end{subequations}
On solving Eqs.~\eqref{P03_eq7}, we obtain a cubic equation in terms of $w_{0}$ is obtained which is given by~\cite{Zhao2008}
\begin{equation}
    w_{0}^3 + \frac{4h^2}{(7-\nu)} w_{0} - \frac{3(1-\nu)Pa^4}{Eh} =0.
    \label{P03_eq8}
\end{equation}
Solving Eq.~\eqref{P03_eq8} using the Cardan's Formula yields the Eq.~\eqref{P03_eq2}.

\section{{Strain along the direction of transport}\label{P03_app3}}
The strain ($\varepsilon$) along a particular direction of transport is different at different portions of the circular membrane, as shown in Fig.~\ref{P03_2}.  Thus, we have to calculate the average strain in the membrane due to applied pressure. We divide the circular basement of the deflected membrane into a large number of parallel chord paths along the direction of transport.  The strain in the geodesic arcs of the deflected membrane along these chord paths are given by the difference in the length of the geodesic arc and the length of the corresponding chords. \\
\indent The length of the chord in an unstrained membrane passing through $(x_{i},0)$ is equal to $L_{0}=\sqrt{a^2-{x_{i}}^2}$. The length of the line passing through the deflected surface is given by
\begin{equation}
    L_{i}= \int\limits_0^{\sqrt{a^2-x_i^2}}\sqrt{1+{\bigg(\frac{dw}{dy}\bigg)^2}}dy.
    \label{P03_eq9}
\end{equation}
The value of strain $\varepsilon_{i}$ can be written as
\begin{equation}
    \varepsilon_{i}= \frac{L_{i}-L_{0}}{L_{0}}.
    \label{P03_eq10}
\end{equation}
On simplification of Eq. C1 using Eq.~B1, we obtain  
\begin{equation}
    L_{i}=\frac{Y\sqrt{1+Y^2}}{2A} + \frac{1}{2A}log|Y+\sqrt{1+Y^2}|,
    \label{P03_eq11}
\end{equation}
where $Y=-\frac{-2w_{0}\sqrt{a^2-{x_i}^2}}{a^2}$ and $A=\frac{-2w_0}{a^2}$. Further, using Eqs.~\eqref{P03_eq10} and \eqref{P03_eq11}, we obtain  the strain along the direction of transport.

\section{{Pressure sensitivity per unit area of graphene membrane} \label{P03_app4}}

The PS of a graphene membrane is given by Eq.~\eqref{P03_eq1}. The GF of graphene is very small across different length scales due to the presence of robust Dirac cones~\cite{Sinha2019, Sinha2020}. According to Eq.~\eqref{P03_eq1}, PS depends on the magnitude of strain at a particular pressure. The radial ($\varepsilon_{r}$) and tangential strains ($\varepsilon_{\theta}$) are expressed as
\begin{subequations} \label {P03_eq12}
\begin{align}
  \varepsilon_{r}&=\frac{3-\nu}{4}\bigg(\frac{w_{0}}{a}\bigg)^{2}\bigg[1-\frac{1-3\nu}{3-\nu}\bigg(\frac{r}{a}\bigg)^{2} \bigg], and \\
  \varepsilon_{r}&=\frac{3-\nu}{4}\bigg(\frac{w_{0}}{a}\bigg)^{2}\bigg[1-\bigg(\frac{r}{a}\bigg)^{2} \bigg]
\end{align}
\end{subequations}
respectively~\cite{Zhao2008}, where $\nu$ is the Poisson's ratio. At any point in the deflected membrane, strain components are directly proportional to $\frac{w^{2}_{0}}{a^2}$. Thus, PS is directly proportional to $\frac{w^{2}_{0}}{a^2}$, and PS per unit area is directly proportional to $\frac{w^{2}_{0}}{a^4}$.

\bibliography{reference01}

\end{document}